\documentclass[12pt,preprint]{aastex}

\newcommand\TA{\tablenotemark{a}}
\newcommand\TB{\tablenotemark{b}}
\newcommand\TC{\tablenotemark{c}}
\newcommand\TD{\tablenotemark{d}}
\newcommand\TE{\tablenotemark{e}}
\newcommand\TF{\tablenotemark{f}}

\newcommand\mcnd{\multicolumn{2}{c}{\nodata}}

\shorttitle{NGC~5307 Spectrophotometry}
\shortauthors{Ruiz et al.}

\received{}
\begin{document}

\title{VLT Echelle Spectrophotometry of the Planetary Nebula NGC~5307
and Temperature Variations
\footnotemark{}}

\author{Mar\'{\i}a Teresa Ruiz\altaffilmark{1}}
\affil{Departamento de Astronom\'{\i}a, Universidad de Chile,
Casilla Postal 36D, Santiago de Chile, Chile}
\email{mtruiz@das.uchile.cl}
   
\author {Antonio Peimbert}
\affil {Instituto de Astronom\'\i a, UNAM,
Apdo. Postal 70-264, M\'exico 04510 D.F., Mexico} 
\email{antonio@astroscu.unam.mx}

\author{Manuel Peimbert}
\affil{Instituto de Astronom\'\i a, UNAM, 
Apdo. Postal 70-264, M\'exico 04510 D.F., Mexico}
\email{peimbert@astroscu.unam.mx}

\and

\author{C\'esar Esteban}
\affil{Instituto de Astrof\'\i sica de Canarias, E-38200 La Laguna, Tenerife, 
Spain}
\email{cel@ll.iac.es}

\begin{abstract}
Echelle spectrophotometry of the planetary nebula NGC~5307 is presented. The
data consists of VLT UVES observations in the 3100 to 10360 \AA\ range.
Electron temperatures and densities have been determined using different line
intensity ratios. We determine the H, He, C, and O abundances based on
recombination lines, these abundances are almost independent of the temperature
structure of the nebula. We also determine the N, O, Ne, S, Cl, and Ar
abundances based on collisionally excited lines, the ratios of these abundances
relative to that of H depend strongly on the temperature structure of the
nebula.  From the \ion{O}{2}/[\ion{O}{3}] ratios we find a $t^2 = 0.056 \pm
0.005$.  The chemical composition of NGC~5307 is compared with those of the Sun
and the Orion nebula. From the study of the relative intensities of the
\ion{O}{2} recombination lines of multiplet 1 in this and other nebulae it is
found that for electron densities smaller than about 5000 cm$^{-3}$ collisional
redistribution is not complete, this effect has to be taken into account to
derive the O abundances for those cases in which not all the lines of the
multiplet are observed. From the $\lambda$ 4649 \ion{O}{2} versus
$N_e$(\ion{Cl}{3}) diagram we find a critical electron density of 1325
cm$^{-3}$ for collisional redistribution of the \ion{O}{2} lines of multiplet
1. Also based on this diagram we argue that the \ion{O}{2} and the [\ion{O}{3}]
lines originate in the same regions.  We also find that the radial velocities
and the FWHM of the \ion{O}{2} and [\ion{O}{3}] lines in NGC~5307 are similar
supporting the previous result.  These two results imply that for NGC 5307 and
probably for many other gaseous
nebulae chemical inhomogeneities are not responsible for the large temperature
fluctuations observed.
\end {abstract}

\keywords{planetary nebulae: abundances---planetary nebulae: individual
(NGC~5307)}

\section{Introduction}

\footnotetext{Based on observations collected at the European Southern
Observatory, Chile, proposal number ESO 68.C-0149(A).}

The main aim of this paper is to make a new determination of the chemical
abundances of NGC~5307 including the following improvements over previous
determinations: the consideration of the temperature structure that affects the
helium and heavy elements abundance determinations, the derivation of the O and
C abundances from recombination line intensities, the consideration of the
collisional excitation of the triplet \ion{He}{1} lines from the $2^3$S level
by determining the electron density from many line intensity ratios, and the
study of the $2^3$S level optical depth effects on the intensity of the triplet
lines by observing a large number of singlet and triplet lines of
\ion{He}{1}. To determine the temperature structure it is also important to
discuss if the \ion{O}{2} and the [\ion{O}{3}] lines originate in the same
regions under the same physical conditions. To this effect we will study the
relative line intensities of the \ion{O}{2} multiplet 1 in this object and
other gaseous nebulae, moreover for the same reason we will also consider the
wavelengths and line widths of the \ion{O}{2} and [\ion{O}{3}] lines in
NGC~5307.

In sections 2 and 3 the observations and the reduction procedure are
described. In section 4 temperatures and densities are derived from five and
four different intensity ratios respectively; also in this section, the mean
square temperature fluctuation, $t^2$, is determined from the
\ion{O}{2}/[\ion{O}{3}] line intensity ratios.  In section 5 ionic abundances
are determined based on recombination lines that are almost independent of the
temperature structure, and ionic abundances based on ratios of collisionally
excited lines to recombination lines that do depend on the temperature
structure of the nebula.  In section 6 the total abundances are determined and
compared with other determinations for the same object. In sections 7 and 8 we
present the discussion and the conclusions.

\section{Observations}

The observations were obtained with the Ultraviolet Visual Echelle
Spectrograph, UVES \citep{dod00}, at the VLT Kueyen Telescope in Chile. We
observed simultaneously with the red and blue arms, each in two settings, covering
the region from 3100 \AA\ to 10360 \AA ~(see Table~\ref{tobs}). The wavelength
regions 5783--5830 \AA\AA\ and 8540--8650 \AA\AA\ were not observed due to the
separation between the two CCDs used in the red arm. There were also two small
gaps that were not observed, 10084--10088 \AA\AA\ and 10252--10259 \AA\AA,
because the two redmost orders did not entirely fit within the CCD.  In addition to the
long exposure spectra we took 60 second exposures for the four observed
wavelength ranges to check for possible saturation effects.

The slit was oriented east-west and the atmospheric dispersion corrector (ADC)
was used to keep the same observed region within the slit regardless of the air
mass value.  The slit width was set to 3.0" and the slit length was set to 10"
in the blue arm and to 12" in the red arm; the slit width was chosen to
maximize the S/N ratio of the emission lines and to maintain the required
resolution to separate most of the weak lines needed for this project. The FWHM
resolution for the NGC~5307 lines at a given wavelength is given by $\Delta
\lambda \sim \lambda / 8800$. The reductions were made for an area of 3"
$\times$ 10" . The center of the slit was placed 5" south of the central star,
a region of high emission measure. The dimensions of the object are
18.8" $\times$ 12.9" \citep[e. g.][]{tyl03}. The average seeing during
the observations amounted to 0.8".

The spectra were reduced using the IRAF\footnotemark{} echelle reduction
package, following the standard procedure of bias subtraction, aperture
extraction, flatfielding, wavelength calibration and flux calibration. For flux
calibration the standard star EG~247 was observed; the observing time
amounted to 600 seconds in each of the four observed wavelength ranges.

\footnotetext{IRAF is distributed by NOAO, which is operated by AURA,
under cooperative agreement with NSF.}

\section{Line Intensities, Reddening Correction and Radial Velocities}

Line intensities were measured integrating all the flux in the line between two
given limits and over a local continuum estimated by eye. In the few cases of
line-blending, the line flux of each individual line was derived from a
multiple Voigt profile fit procedure. All these measurements were carried out
with the {\tt splot} task of the IRAF package.

The reddening coefficient, $C$(H$\beta$) = 0.59 $\pm 0.10$ dex, was determined
by fitting the observed $I$(H$\beta$)/$I$(H Balmer lines) ratios to the
theoretical ones computed by \citet{sto95} for $T_e$ = 10,000 K and $N_e$ =
2000 cm$^{-3}$, (see below) and assuming the extinction law of \citet{sea79}.

Table~\ref{tlines} presents the emission line intensities of NGC~5307.  The
first and second columns include the observed wavelength in the heliocentric 
framework, $\lambda$, and the adopted laboratory wavelength,
$\lambda_0$. The third and fourth columns include the ion and the multiplet
number, or the Balmer, B, or Paschen, P, transitions for each line. The main 
sources used for the line identifications and laboratory wavelenghts were: 
\citet{est98,hyu00,liu01,wie96,peq88}, and references therein.  The fifth
and sixth columns include the observed flux relative to H$\beta$, $F(\lambda$),
and the flux corrected for reddening relative to to H$\beta$, $I(\lambda$). To
combine all the line intensities, from the four different instrumental settings, 
on the same scale, we multiplied 
the intensities in each setting by a correction factor obtained from the lines 
present in more than one setting; these corrections were 
all smaller than 2\%. The seventh column includes the 
fractional error (1$\sigma$) in the line intensities; these errors were 
estimated from a signal-to-noise versus signal curve obtained from a fit to the 
dispersion in the observed to predicted line intensities for all the measured 
hydrogen lines.

A total of 184 distinct emission lines were measured; of them 137 are
permitted, 46 are forbidden and 1 is semiforbidden (see Table~\ref{tlines}). In
this tally only the main component of the blended lines is included. Many lines 
redward of 7000 \AA\ inclunding several Paschen lines and 
$\lambda$ 9531 [\ion{S}{3}] were detected, but
they were strongly affected by atmospheric features in absorption and emission
rendering their intensities unreliable.

We derived the heliocentric radial velocity of NGC~5307 based on the 14
strongest \ion{H}{1} and \ion{He}{1} lines and amounts to $+ 29.4 \pm 0.1$ km
sec$^{-1}$.  This value is smaller than that derived by \citet{mea88} that
amounts to $+ 40 \pm 4$ km sec$^{-1}$, the difference probably is due to the
different areas observed by the two groups.

The center of our observing slit was located at about half the distance
between the central star and the outer edge of the main body of the nebula.
Most of the emission lines show a single line profile because we are
observing material moving from a tangential velocity up to an angle of 60
degrees relative to the plane of the sky. The exceptions are the blue and 
red components of the $\lambda\lambda$ 6313 [\ion{S}{3}], 6716
[\ion{S}{2}], 6731 [\ion{S}{2}], and 9069 [\ion{S}{3}] lines that show a
splitting of $35 \pm 3$ km sec$^{-1}$, because they mainly originate in the outer
regions of the nebula. Dividing this value by the sine of 60 degrees we
find a 2$V_{exp}$ value of about 40 km sec$^{-1}$ of this material relative
to the central star. This value is similar to the 2$V_{exp}$ values of 
30 km sec$^{-1}$ and 21.2 km sec$^{-1}$ found by \citet{ack76} and \citet{mea88} 
respectively.

\section{Physical Conditions}

\subsection{Temperatures and densities}

The temperatures and densities presented in Table~\ref{ttemden} were derived
from the line intensities presented in Table~\ref{tlines}. Most of the
determinations were carried out based on the IRAF subroutines.

The contribution to the intensities of the $\lambda\lambda$ 7319, 7320, 7331,
and 7332 [\ion{O}{2}] lines due to recombination was taken into account based
on the following equation:
\begin{equation}
I_R(7319+7320+7331+7332)/I({\rm H\beta})
= 9.36(T/10^4)^{0.44} \times {\rm{O}}^{++}/{\rm{H}}^+,
\end{equation}
\citep[see][]{liu00}.  

The contribution to the intensity of the $\lambda$ 5755 [\ion{N}{2}] line due
to recombination was taken into account based on the following equation:
\begin{equation}
I_R(5755)/I({\rm H\beta})
= 3.19(T/10^4)^{0.30} \times {\rm{N}}^{++}/{\rm{H}}^+,
\end{equation}
\citep[see][]{liu00}.  

The Balmer continuum temperature was determined from the following equation:
\begin{equation}
T = 368 \times(1 + 0.259y^+ + 3.409y^{++})({\rm Bac/H11})^{-3/2} ~{\rm {K}},
\end{equation}
\citep[see][]{liu01} where Bac/H11 is in \AA$^{-1}$, $y^+$ = He$^+$/H$^+$, and
$y^{++}$ = He$^{++}$/H$^+$. 

\subsection{Temperature variations}

To derive the ionic abundance ratios the average temperature, $T_0$, and the
mean square temperature fluctuation, $t^2$, were used. These quantities are
given by
\begin{equation}
T_0 (N_e, N_i) = \frac{\int T_e({\bf r}) N_e({\bf r}) N_i({\bf r}) dV}
{\int N_e({\bf r}) N_i({\bf r}) dV},
\end{equation}
and
\begin{equation}
t^2 = \frac{\int (T_e - T_0)^2 N_e N_i dV}{T_0^2 \int N_e N_i dV},
\end{equation}
respectively, where $N_e$ and $N_i$ are the electron and the ion densities of
the observed emission line and $V$ is the observed volume \citep{pei67}.

To determine $T_0$ and $t^2$ we need two different methods to derive $T_e$: one
that weighs preferentially the high temperature regions and one that weighs
preferentially the low temperature regions. In this paper we have used the
temperature derived from the ratio of the [\ion{O}{3}] $\lambda\lambda$ 4363,
5007 lines, $T_{(4363/5007)}$, that is given by
\begin{equation}
T_{(4363/5007)} = T_0 \left[ 1 + {\frac{1}{2}}\left({\frac{90800}{T_0}} - 3
\right) t^2\right],
\end{equation}
and the temperature derived from the ratio of the recombination lines of
multiplet 1 of \ion{O}{2} to the collisionally excited lines of [\ion{O}{3}]
that is given by

\begin{equation}
T_{({\rm O\,\scriptscriptstyle II}\,rec /
{\rm O\,\scriptscriptstyle III}\,coll)} =
T_{(4649/5007)} = f_1(T_0,t^2).
\end{equation}

{From} these equations we obtain that for O$^{++}$(R/C): $t^2$ = 0.056 $\pm
0.005$, and $T_0$ = 10100 K. Since about 90 $\%$ of the oxygen is twice ionized
(see below) we will use this $t^2$ value as representative for the whole
object. From $T$(Bac), $T$(\ion{O}{2}), $T$(\ion{O}{3}), and equations
6-9 by \citet{pei00} we obtain $t^2=0.03 \pm 0.05$ this value is not accurate 
enough to be useful in the abundance determinations.

\section{Ionic Chemical Abundances}

\subsection{Helium ionic abundances}

To obtain He$^+$/H$^+$ values we need a set of effective recombination
coefficients for the He and H lines, the contribution due to collisional
excitation to the helium line intensities, and an estimate of the optical depth
effects for the helium lines. The recombination coefficients used were those by
\citet{sto95} for H, and those by \citet{smi96} and \citet{ben99} for He. The
collisional contribution was estimated from \citet{saw93} and
\citet{kin95}. The optical depth effects in the triplet lines were estimated
from the computations by \citet*{ben02}.

Table~\ref{thelium} presents the He$^+$/H$^+$ values obtained from the eight
best observed helium lines for $N_e = 2500 \pm 500$. From a maximum likelihood
method we obtained He$^+$/H$^+ = 0.09055 \pm 0.00151$, $\tau_{3889} = 0.93 \pm
0.57$, and $t^2 = 0.031 \pm 0.014$

{From} the $I$(4686)/$I$(H$\beta$) value presented in Table~\ref{tlines}, $T_e$
= 13500 K, and the recombination coefficients by \citet{bro71} we obtain
$N$(He$^{++}$)/$N$(H$^+) = 0.00726 \pm 0.00018$. The helium ionic abundances
are presented in Table~\ref{trl}.

\subsection{C and O ionic abundances from recombination lines}

The C$^{++}$ abundance was derived from the $\lambda 4267$ \AA~line of
\ion{C}{2} and the effective recombination coefficients computed by
\citet{dav00} for Case A and $T$ = 10000 K.

The O$^{++}$ abundance was derived from the lines of multiplet 1 of
\ion{O}{2} (see Figure~\ref{foxygen}) together with the effective recombination
coefficient for the multiplet computed by \citet{sto94} under the assumption of
Case B for $T_e = 10000$ K and $N_e$ = 2000 cm$^{-3}$.

Six  of the eight lines of multiplet 1  were observed and are presented in 
Table~\ref{tlines}, the $\lambda\lambda$ 4674 and 4696 \AA\ lines were too 
weak to be observed, we computed their intensities as follows:
a I(4674)=0.019 value was determined from I(4651) and the ratio of their 
Einstein A coefficients, since both lines originate from the same upper level; 
similarly a I(4696)=0.008 value was determined from I(4639) + I(4661) and 
the ratio of their A coefficients, since the 3 lines originate from the same 
upper level. The A values were taken from \citet{wie96}. 
Additionally the $\lambda 4641.81$ \AA~line of \ion{O}{2} is blended with 
the $\lambda 4641.85$ \AA~line of \ion{N}{3}, based on the observed intensities 
and the A values of other lines of the same multiplets originating in the same 
upper energy levels we estimated that the intensities of these lines, in the 
units of Table~\ref{tlines}, amount to 0.148 and 0.053 respectively. 

The O$^{++}$ abundance is almost independent
of the case assumed, the difference in the O$^{++}$/H$^+$ value between Case A
and Case B is smaller than 4\%. \citet{pea03} found that in 30 Doradus the
\ion{O}{2} lines of multiplet 1 are in Case B based on the observed intensities
of multiplets 19, 2, and 28 of \ion{O}{2} that are strongly case sensitive;
\citet{pei93y} also found that the \ion{O}{2} lines in the Orion nebula are in
Case B. The line intensity ratios within multiplet 1 do not follow the 
predictions for collisional redistribution, this result will be discussed in section 7.
Figure~\ref{foxygen} provides an excellent visual reference to estimate the
quality of the data, it includes 2 lines four orders of magnitude fainter than
H$\beta$ and also shows that lines separated by 2\AA\ are completely resolved.

The C$^{+3}$ abundance was derived from: a) $\lambda 4647$ \AA~, one of the
three lines of multiplet 1 of \ion{C}{3}, b) the effective recombination
coefficient used by \citet{liu00} for the whole multiplet, and c) the
assumption that $I$(4647 + 4650 + 4651)/$I$(4647) = 1.8, the LTE ratio
for multiplet 1 \citep{wie96}.

Most of the intensity of each of the \ion{O}{3} and \ion{N}{3} lines 
present in Table~\ref{tlines} is due to the well known Bowen resonance 
fluorescence mechanism \citep{bow34} making these lines and are not 
suitable for accurate abundance determinations.
 
\subsection{Ionic abundances from collisionally excited lines}

The values presented in Table~\ref{tcl} for $t^2 = 0.00$ were derived with the
IRAF task {\tt abund}. For \ion{N}{2}, \ion{O}{2}, and \ion{S}{2} we used $T_e$
= 11000 K and $N_e$ = 3500 cm$^{-3}$.  For \ion{O}{3}, \ion{Ne}{3}, \ion{S}{3},
\ion{Cl}{3}, \ion{Cl}{4}, \ion{Ar}{3}, and \ion{Ar}{4} we used $T_e$ = 11800 K
and $N_e$ = 2500 cm$^{-3}$. For \ion{Ar}{5} and \ion{K}{4} we used $T_e$ =
13500 K and $N_e$ = 2500 cm$^{-3}$.

To derive the abundances for $t^2 = 0.056$ we used the abundances for $t^2 =
0.00$ and the following equations for $t^2 > 0.00$ presented by \citet{pei67}
and \citet{pei69}:
\begin{equation}
\left[ {N({\rm CEL}) \over N({\rm RL})} \right]_{true} = 
{{T_{\rm RL}}^{\alpha} \ {T_{\rm CEL}}^{0.5} \over {T_{(4363/5007)}}^{\alpha + 0.5}} \times
\exp  \left( -{\Delta E_{\rm CEL} \over kT_{(4363/5007)}} +{\Delta E_{\rm CEL} \over kT_{\rm CEL}} \right)
\times \left[ {N({\rm CEL}) \over N({\rm RL})} \right]_{(4363/5007)},
\label{ecelrl}
\end{equation}
\begin{equation}
T_{\rm CEL} = T_0 \left\{ 1 + \left[ 
{\left( \Delta E_{\rm CEL}/kT_0 \right)^2 - 3\Delta E_{\rm CEL}/kT_0 + 3/4 
\over \Delta E_{\rm CEL}/kT_0 - 1/2}
\right] {t^2 \over 2} \right\},
\end{equation}
\begin{equation}
T_{\rm RL} = T_0 \left[ 1 + \left( 1 - \alpha \right) {t^2 \over 2} \right],
\end{equation}
\begin{equation}
T_{(4363/5007)} = 
T_0 \left[ 1 + \left( {90800 \over T_0} - 3 \right) {t^2 \over 2} \right],
\end{equation}
where $\alpha$ is the dependence on the temperature for a given recombination 
line, $\Delta E_{\rm CEL}$ is the energy difference between the ground and the
excited level, and [$N$(CEL)/$N$(RL)]$_{(4363/5007)}$ is the abundance ratio
based on the temperature obtained from $I$(4363)/$I$(5007), note that the
temperature exponents presented by \citet{pei69} for equation~(\ref{ecelrl})
are misprinted.

To derive abundances for other $t^2$ values it is possible to interpolate or to
extrapolate from the values presented in Table~\ref{tcl}.

\section{Total Abundances}

Table~\ref{tabund} presents the total gaseous abundances of NGC~5307 for
$t^2=0$ and $t^2 = 0.056$.  To derive the total gaseous abundances the set of
equations presented below was used, where the ionization correction factors,
$ICF$'s, correct for the unseen ionization stages.

The total He/H value is given by:
\begin{equation}
\nonumber
\frac{N ({\rm He})}{N ({\rm H})} =  \frac {N({\rm He}^+) + N({\rm He}^{++})}
{N({\rm H}^+)}.                 
\end{equation}

The C abundance is given by:
\begin{equation}
\frac{N(\rm C)}{N(\rm H)} = ICF({\rm C}) \frac{N({\rm C^{++}}) + N({\rm C^{+3}})}{N(\rm H^+)}.
\end{equation}
The $ICF$(C) was estimated from models computed with CLOUDY \citep{fer96,fer98}
with similar O and He ionization structure and amounts to 1.05, a value very close to 1. 
Half of the correction is due to the expected C$^+$ contribution and half to the C$^{+4}$
contribution.

The following equation has been used often to obtain the N/H abundance ratio
\citep{pei69}:
\begin{equation}
\frac{N(\rm N)}{N(\rm H)} = ICF({\rm N}) \frac{N({\rm N^+})}{N({\rm H^+})} = 
              \frac{N({\rm O})}{N({\rm O^+})}
              \frac{N({\rm N^+})}{N({\rm H^+})}.
\label{eN1}
\end{equation}
{From} our observations of NGC~5307, $ICF$(N) is in the 30 to 40 range, a very
large and consequently uncertain correction. \citet{rel02,lur02} have obtained
$ICF$(N$^+$) values bigger than those given by equation~(\ref{eN1}). Therefore we
decided to use a different equation to obtain N/H given by:
\begin{equation}
 \frac{N({\rm N})}{N({\rm H})} =
           \left( \frac{N({\rm O})}{N({\rm H^+})} \right)_{V}
           \left( \frac{N({\rm N}^{++}) + N({\rm N}^{+3})}{N({\rm O}^{++}) + N({\rm O}^{+3})} \right)_{UV};           
\label{eN2}
\end{equation}
where the first term on the right hand side was obtained from
Table~\ref{tabund} and the N and O ionic abundances were obtained from: a) the
line intensities of $\lambda\lambda$ 1401 \ion{O}{4}, 1486 \ion{N}{4}], 1660 +
1666 \ion{O}{3}], and 1750 \ion{N}{3}] presented by \citet{kin94}, b) the
atomic parameters presented in the compilation by \citet{men83} for
$\lambda\lambda$ 1660 + 1666 \ion{O}{3}], and the atomic parameters in IRAF for
the other lines, and c) a $T_e$ of 12500 K. The values derived from
equation~(\ref{eN2}) are those presented in Table~\ref{tabund}.

We prefer equation~(\ref{eN2}) to equation~(\ref{eN1}) because about
97~-~99$\%$ of the N an O atoms are expected to be doubly or triply ionized,
while only about 1~-~3 $\%$ of the N and O atoms are expected to be singly
ionized.

It is also possible to determine N/H from:
\begin{equation}
\nonumber
\frac{N ({\rm N })}{N ({\rm H})} =  \frac {[N({\rm N}^{++}) + N({\rm N}^{+3})]_{UV}}
{[N({\rm H}^+)]_V};                 
\label{eN3}
\end{equation}
we prefer equation~(\ref{eN2}) to equation~(\ref{eN3}) because there is a
mismatch between the abundances derived from $UV/V$ line intensity ratios and
those derived from the $UV/UV$ or $V/V$ ratios. For example, from the $UV$ data
by \citet{kin94} and their $I$(H$\beta$) value we obtain
$N$(O)$_{UV}$/$N$(H)$_V = 12.65 \times 10^{-4}$ for $t^2$ = 0.00, while from
their visual data we obtain $N$(O)$_V$/$N$(H)$_V = 4.3 \times 10^{-4}$, also
for $t^2$ = 0.00, a difference of a factor of three. This difference could be
due to errors in the $C$(H$\beta$) value, errors in the adopted temperature,
and differences in the slit size and position between the $UV$ and $V$
observations. Alternatively we consider the [$N$(N)/$N$(O)]$_{UV}$ value to be
more reliable than the $N$(N)$_{UV}$/$N$(H)$_V$ value given by
equation~(\ref{eN3}), because for the $UV$ line ratios the dependence on
$C$(H$\beta$) and $T_e$ is smaller and the slit size and position are the same.
Similarly we consider [$N$(O)/$N$(H)]$_V$ to be more reliable than
$N$(O)$_{UV}$/$N$(H)$_V$ because the visual line ratios depend less on
$C$(H$\beta$) and $T_e$ than the $UV$ to $V$ line ratios, and the slit size and
position are the same for the visual line intensities.

To derive the same N/H value from equations (\ref{eN1}) and~(\ref{eN2}) we need
an $ICF$(N) twice as large as that provided by equation~(\ref{eN1}). A similar
increase in the $ICF$(N) value was obtained for NGC~346, the brightest
\ion{H}{2} region in the SMC, by \citet{rel02} based on CLOUDY ionization
structure models.

The O abundance was obtained from \citep{kin94}:
\begin{eqnarray}
\nonumber
\frac{N({\rm O})}{N({\rm H})} & = & 
           ICF({\rm O})
           \left( \frac{N({\rm O^+})+N(\rm O^{++})}{N(\rm H^+)} \right) \\
                               & = &
           \left( \frac{N({\rm He}^+) + N({\rm He}^{++})}{N({\rm He}^+)} \right)^{2/3} 
           \left( \frac{N({\rm O^+})+N(\rm O^{++})}{N(\rm H^+)} \right),
\end{eqnarray}
where $ICF$(O) takes into account the fraction of O$^{+3}$.

The Ne abundance was obtained from \citep{pei69}:
\begin{eqnarray}
\nonumber
\frac{N({\rm Ne})}{N({\rm H})} & = &
             ICF({\rm Ne})       
             \left( \frac{N({\rm Ne}^{++})}{N({\rm H}^+)} \right) \\
                               & = & 
             \left( \frac{N({\rm O})}{N({\rm O}^{++})} \right)
             \left( \frac{N({\rm Ne}^{++})}{N({\rm H}^+)} \right).
\end{eqnarray}

The S abundance was obtained from:
\begin{equation}
\label{esulphur}
\frac{N({\rm} S)}{N({\rm H})} = ICF({\rm S}) \frac{N({\rm S^+}) + N({\rm
S}^{++})}{N({\rm H}^+)},
\end{equation}
where $ICF$(S) was estimated from the models by \citet{gar89} and amounts to
$6.3 \pm 1.5$.

The Cl abundance is given by:
\begin{equation}
\frac{N(\rm Cl)}{N(\rm H)} = ICF({\rm Cl})\frac{N({\rm Cl^{++}}) + N(\rm Cl^{+3})}{N(\rm H^+)},
\end{equation}
where $ICF$(Cl) was estimated from models computed with CLOUDY \citep{fer96,fer98}
with similar O and He ionization structure and amounts to 1.05, a value very close to 1.
the correction is due  Cl$^{+4}$, the amount of Cl$^{+}$ for this object is negligible.

Finally the Ar abundance is given by:
\begin{equation}
\frac{N(\rm Ar)}{N(\rm H)} = ICF({\rm Ar}) \frac{N({\rm Ar^{++}}) + 
N(\rm Ar^{+3}) + N(\rm Ar^{+4})}{N(\rm H^+)},
\end{equation}
where $ICF$(Ar) = 1/[1 - O$^+$/O] = 1.02 \citep{liu00}, takes into account the
Ar$^{+}$ fraction.

\section{Discussion}

\subsection{The $\lambda$ 4649 \ion{O}{2} versus electron density diagram}

The relative intensity ratios of the \ion{O}{2} recombination lines of
multiplet 1 do not agree with the LTE computation predictions,
this deviation was noted before for other objects by \citet{est99a};
\citet{est99b}; \citet{pea03}; and \citet{tsa03}. In particular \citet{tsa03}
discuss this problem extensively.  

To explain this deviation, we propose that the electron density is smaller 
than that needed to produce an LTE collisional redistribution of the fine 
structure energy levels. To test our proposal we produced Figure~\ref{fOvsden} 
that is based on the best \ion{O}{2} observations in the literature, those 
presented in Table~\ref{toii}. We decided to plot $I$(4649) over the sum of 
the intensities of all the lines of the multiplet $I$(sum), versus the density 
derived from optical forbidden line ratios, mainly the [\ion{Cl}{3}] lines 
that originate in the same regions where O is twice ionized.

{F}rom Figure~\ref{fOvsden} we find an excellent correlation between
$I$(4649)/$I$(sum) and the forbidden line electron density, 
$N_e$(FL). We propose to adjust the data with a curve given by
\begin{equation}
\left[ \frac{I(4649)}{I({\rm sum})} \right]_{obs} = 
\left[ \frac{I(4649)}{I({\rm sum})} \right]_{cas} + 
\frac{ \left[ \frac{I(4649)}{I({\rm sum})} \right]_{LTE} - 
       \left[ \frac{I(4649)}{I({\rm sum})} \right]_{cas} }
     { \left[ 1 + \frac{N_e({\rm crit})}{N_e({\rm FL})} \right]},
\label{ecurve1}
\end{equation}
where $[I(4649)/I(\rm sum)]_{cas}$ is the value produced by recombination and
cascade without collisional redistribution, $[I(4649)/I(\rm sum)]_{LTE}$
amounts to 0.397 \citet{wie96} the value predicted by collisional
redistribution assuming that the populations of the upper energy levels of
multiplet 1 are given by the statistical weights multiplied by the Einstein A
values, and $N_e$(crit) is the critical density defined by
equation~(\ref{ecurve1}). From Figure~\ref{fOvsden} we find $N_e({\rm crit}) = 
1325 \pm 275$ cm$^{-3}$, and $I(4649)/I$(sum)]$_{cas} = 0.105 \pm 0.015$, 
therefore equation~(\ref{ecurve1}) can be written as:
\begin{equation}
\left[ \frac{I(4649)}{I({\rm sum})} \right]_{obs} = 
0.105 + \frac{0.292}{ \left[ 1 + 1325/N_e({\rm FL}) \right] }. 
\label{ecurve2}
\end{equation}
It is beyond the scope of this paper to check this equation based on an atomic
physics computation.

Equation~\ref{ecurve1} has the following applications:
a) when $N_e({\rm FL})$ is known it is possible to use this equation to 
obtain $I$(sum) from the observed $I(4649)$ and from $I$(sum) to obtain 
the O$^{++}$ abundance since $I$(sum) is independent the density, from 
$I(4649)$ alone it is not possible to determine an accurate O$^{++}$ 
abundance since this line is density dependent; b) from an observed 
$I$(4649)/$I$(Sum) it is possible to determine $N_e$ in the O$^{++}$ zone; 
c) if the forbidden line density of a nebula does not fall close to the 
value predicted with this equation it could mean 
that there are extreme inhomogeneities in density or chemical 
composition in this object, for example an object with high-density 
oxygen-rich knots embedded in a low density medium would fall to the 
left of the curve.

From Table~\ref{toii} it can be seen that the intensity of the lines that originate 
in the 3p~$^4$D$^0_{3/2}$ and 3p $^4$D$^0_{1/2}$ energy levels decreases with 
increasing local density, while the intensity of those lines that originate in the 
3p~$^4$D$^0_{7/2}$ increases with increasing density. This is due to the effect 
of collisional redistribution that increases the population of the high 
statistical weight levels at the expense of the low statistical weight ones. 
Note that the intensity of the lines that originate from the 3p~$^4$D$^0_{5/2}$ 
energy level are almost independent of the electron density.

The results derived from Figure~\ref{fOvsden} 
imply that in these objects the [\ion{O}{3}] $\lambda\lambda$ 4363, 4959, and
5007 lines and the \ion{O}{2} recombination lines of multiplet 1 originate in
regions of similar density, this is particularly the case for those objects
with $N_e$(FL) $<$ 3000 cm$^{-3}$. In other words: we can not assume the 
presence of high density knots to explain the difference between the O$^{++}$
abundances derived from the [\ion{O}{3}] lines and the \ion{O}{2} lines.
Consequently we propose to drop the constant temperature assumption, and to 
explain the apparent difference in abundances as being due to the presence 
of temperature fluctuations.

In general to reconcile the abundances derived from collisionally excited lines
and recombination lines it is necessary to assume the presence of large
temperature variations. The presence of temperature variations in gaseous
nebulae has been amply reviewed in the recent literature
\citep[e.g.][]{pei02,est02a,liu02,tor03}. Even more recent results in favor of
large temperature variations in the Orion nebula, the 30 Doradus nebula, and
other planetary nebulae have been presented by \citet{rub03,ode03,pea03,pei03}.

\citet{pei93y}, by comparing the \ion{O}{2} line intensities of multiplets 1, 2, 
and 10 for for the planetary nebula NGC~6572, have concluded that the \ion{O}{2} 
lines are produced by recombination, and that the contribution of resonance 
fluorescence to the \ion{O}{2} line intensities is not important. A similar 
conclusion was reached \citet{gra76} for the Orion nebula.

\subsection{The heliocentric velocities and the FWHM of the \ion{O}{2} and 
[\ion{O}{3}] lines}

The difference between the O(CEL) abundance and the O(RL) abundance indicates
the presence of large temperature variations. There are two possible 
scenarios for these temperature variations: a) that the abundances are 
homogeneous along the line of sight and temperature fluctuations of $\pm 23\%$ are 
present, an explanation for these large temperature 
variations has to be sought \citep[e.g.][and references therein]{tor03}, or 
b) that the abundances are not homogeneous along the line of sight and that 
there are high density knots with excess heavy element abundances embedded 
in a medium with lower heavy element abundances and higher electron temperature
\citep[e.g.][]{jac83,tor90,gue96,peq02,wes03}.

The second possibility is based on the idea that the central stars, of at least
some planetary nebulae, firstly eject an envelope at relatively low velocity
with almost normal heavy element abundances, and later on eject knots or clumps
of material at higher velocities with an excess of heavy element abundances. In
Abell~58 \citet{gue96} studied a hydrogen deficient high velocity knot
blueshifted relative to the main body of the nebula by 120 km s$^{-1}$. while
\citet{wes03} studied three knots in Abell~30, where knot J1 is redshifted
relative to knot J3 by about 45 km s$^{-1}$ and knot J3 is redshifted relative
to knot J4 by about 100 km s$^{-1}$.

By assuming that chemical inhomogeneities are responsible for the temperature
variations in NGC~5307 we would expect most of the flux of the [\ion{O}{3}]
lines to originate in the lower density medium that expands at lower velocity,
and a considerable fraction of the flux of the \ion{O}{2} lines to originate
from knots or clumps that move at higher velocities. Therefore if this were the
case the radial velocity derived from the central wavelength of the 
[\ion{O}{3}] lines could be different from that derived from the \ion{O}{2} 
lines; furthermore the full width at half maximum, FWHM,
of the [\ion{O}{3}] lines would be smaller than that of the \ion{O}{2} lines.

For NGC~5307 the heliocentric velocity of the $\lambda\lambda$ 4363, 4959, 5007
[\ion{O}{3}] lines is 30.1 $\pm 0.5$ km s$^{-1}$, while that of the \ion{O}{2}
lines of multiplet 1 is 33.1 $\pm 4$ km s$^{-1}$. The FWHM of the
$\lambda\lambda$ 4363, 4959, 5007 [\ion{O}{3}] lines is 48.4 $\pm 0.7$ km
s$^{-1}$, while that of the $\lambda$ 4649 \ion{O}{2} line, the best observed
line of multiplet 1, is 44.3 $\pm 7$ km s$^{-1}$. These results imply that
there is no evidence in favor of high velocity O-rich clumps in our
observations of NGC~5307.

\subsection{Comparison with other abundance determinations}

Table~\ref{tabund} presents a comparison between the abundance determinations
of this paper and those by other authors. A discussion of the uncertainties in
the determinations by other authors is beyond the scope of this paper, the
errors of this work are presented in Tables \ref{tcl} and~\ref{tta}. The
average $N$(He)/$N$(H) value derived by the other four groups amounts
to -1.02 dex in excellent agreement with our value of -1.01 dex. The C
abundance derived by us is based on the $\lambda$ 4267 \ion{C}{2} 
recombination line, while those derived by other authors are based on 
UV lines, we consider our determination to be more accurate because it 
is almost independent of the adopted temperature while the abundance 
derived from the UV collisionally excited lines depends strongly on the 
adopted electron temperature (although the \ion{C}{2} recombination 
abundances are still regarded as controversial by other authors). 
The N/H value has been already amply discussed and most of the difference 
among the various groups depends on the use of $V$ or $UV$ lines to 
determine the N abundance. Most of the difference among the O and Ne 
abundances is due to the adopted $t^2$ value.

To discuss the effect of stellar evolution on the observed abundances
in NGC~5307 we present in Table~\ref{tta} the Orion nebula and solar
abundances as representative of the present day interstellar medium
and of the interstellar medium at the time the Sun was formed.
Table~\ref{tta} presents  the solar photospheric values for C,
N, O, Ne, and Ar, and the solar abundances derived from meteoritic data for S,
Cl, and Fe. For the solar initial helium abundance the $Y_0$ by \citet{chr98}
was adopted, and not the photospheric one because, apparently, it has been
affected by settling. Notice that the solar oxygen abundance used in this paper
is the average value of the recent results by \citet*{all01} and \citet{hol01}
that indicate an O/H value of 8.71 dex, a value lower than that derived by
\citet{gre98}. We are also using the C/H value by \citet*{all02}, the N/H and
Ne/H solar abundances were taken from \citet{hol01}, while the S/H, Cl/H, and
Ar/H come from \citet{gre98}, the first two from the meteoritic data, while the
last one from the photospheric data.

The relative abundances of H, He, C, N, and O imply that NGC 5307
is a PN of Type II with a main sequence progenitor star
in the 1.2 to 2.4 solar mass range \citep{pei78,pei90,kin94}.

In Table~\ref{tta} we present the abundances of NGC 5307, the Orion 
nebula (representative of the interstellar medium today), and the Sun 
(expected to be similar to the interstellar abundances when the 
parental star of NGC 5307 was formed). By comparing these abundances 
we reach the following
conclusions: a) the He abundance has not been significantly enriched
by stellar evolution, b) the C abundance has decreased by about
a factor of three, and c) the N abundance has increased by about
a factor of three. These three results taken together indicate that
the nebular material has only gone through part of the CNO cycle,
converting C into N, but without producing significant amounts of
helium. Evolutionary models of intermediate mass stars predict the 
N increase and the C decrease in their outer layers
while the O and He abundances are almost unaffected 
\citep[e. g.][]{tor71}; N increases at the expense of C
because the C/N ratio in the interstellar medium and in the Sun is 
considerably larger than the equilibrium value produced by the CNO cycle 
\citep[e. g.][]{pag97}.

Objects that show O-rich knots, like A30 and A58, show He/H excesses
that are not present in NGC 5307 \citep{gue96,wes03}. Consequently
the chemical composition of NGC 5307 is also against the presence
of O-rich knots in this object 

\section{Conclusions}

NGC~5307 is a typical Type II PN with $t^2 = 0.056 \pm 0.005$. Its expansion
velocity is low, about 15 km s$^{-1}$. Its chemical composition indicates 
that nuclear processing in the parental star of the ejected shell
has not gone beyond the partial conversion of C into N. In particular it has 
a very similar He/H ratio to that present in the interstellar medium at the time
its progenitor star formed

Its chemical composition, together with the FWHM and the radial
velocities of the  \ion{O}{2} lines and  the [\ion{O}{3}] lines, indicate that 
NGC 5307 does not have high density O-rich knots embedded in a low density 
O-normal medium. Consequently the difference between the \ion{O}{2} and 
[\ion{O}{3}] line intensities has to be due to a different cause than chemical 
inhomogeneities.

The comparison of the FWHM of the \ion{O}{2} and [\ion{O}{3}] lines
provides an excellent method to detect the presence O-rich knots, with a high 
electron density and a high velocity relative to the parental star, embedded 
in a low density O-normal medium.

From the $I$(4649)/$I$(sum) versus $N_e$(Cl III) diagram we have found
an equation that provides a good fit to most of the well observed
gaseous nebulae. This fit to the data indicates that most, if not all,
of the \ion{O}{2} recombination lines intensity originates in the same 
regions than the [\ion{O}{3}] collisional lines intensity, and that 
high density O-rich knots are not the main cause responsible, in these 
objects, for the presence of temperature variations.

We acknowledge the referee for a careful reading of the manuscript.
MTR received partial support from FONDAP(15010003), a Guggenheim Fellowship and
Fondecyt(1010404). MP received partial support from DGAPA UNAM (grant IN
114601).

\clearpage

\begin{deluxetable}{l@{\hspace{48pt}}r@{--}lc} 
\tablecaption{Journal of Observations
\label{tobs}}
\tablewidth{0pt}
\tablehead{
\colhead{Date} & 
\multicolumn{2}{c}{$\lambda$ (\AA)} &
\colhead{Exp. Time (s)}}
\startdata
2002 March 12 & 3100 &3880  &  3$\times$90 \\
2002 March 12 & 3730 &4990  &  3$\times$180 \\
2002 March 12 & 4760 &6840  &  3$\times$90 \\
2002 March 12 & 6600 &10360 &  3$\times$180 \\
\enddata
\end{deluxetable} 

\clearpage

\begin{deluxetable}{r@{}lr@{}ll@{ }rr@{}lr@{}lr@{}l} 
\tablecaption{Line Intensities
\label{tlines}}
\tablewidth{0pt}
\tablehead{
\multicolumn{2}{c}{$\lambda$} &
\multicolumn{2}{c}{$\lambda_0$} &
\colhead{Ion} &
\colhead{Id.} &
\multicolumn{2}{c}{$F$\TA} &
\multicolumn{2}{c}{$I$\TB} &
\multicolumn{2}{c}{ err(\%)} }
\startdata
 3133&.128& 3132&.79    & ~\ion{O }{3} & 12      &   7&.12  &  13&.1   &  5&  \\
 3188&.075& 3187&.74    & ~\ion{He}{1} & 7       &   2&.05  &   3&.68  &  8&  \\
 3203&.469& 3203&.10    & ~\ion{He}{2} & 3.5     &   1&.69  &   3&.00  & 10&  \\
 3299&.807& 3299&.36    & ~\ion{O }{3} & 3       &   0&.482 &   0&.814 & 15&  \\
 3312&.673& 3312&.30    & ~\ion{O }{3} & 3       &   1&.03  &   1&.73  & 10&  \\
 3341&.103& 3340&.74    & ~\ion{O }{3} & 3       &   1&.71  &   2&.83  &  8&  \\
 3428&.957& 3428&.65    & ~\ion{O }{3} & 15      &   0&.352 &   0&.559 & 15&  \\
 3444&.406& 3444&.07    & ~\ion{O }{3} & 15      &   2&.78  &   4&.38  &  6&  \\
 3634&.702& 3634&.25    & ~\ion{He}{1} & 28      &   0&.211 &   0&.308 & 20&  \\
 3663&.821& 3663&.40    & ~\ion{H }{1} & H29     &   0&.153 &   0&.222 & 20&  \\
 3664&.943& 3664&.68    & ~\ion{H }{1} & H28     &   0&.234 &   0&.339 & 15&  \\
 3666&.343& 3666&.10    & ~\ion{H }{1} & H27     &   0&.318 &   0&.460 & 15&  \\
 3668&.015& 3667&.68    & ~\ion{H }{1} & H26     &   0&.332 &   0&.480 & 15&  \\
 3669&.823& 3669&.46    & ~\ion{H }{1} & H25     &   0&.382 &   0&.552 & 12&  \\
 3671&.785& 3671&.48    & ~\ion{H }{1} & H24     &   0&.382 &   0&.551 & 12&  \\
 3674&.123& 3673&.76    & ~\ion{H }{1} & H23     &   0&.460 &   0&.664 & 12&  \\
 3676&.712& 3676&.36    & ~\ion{H }{1} & H22     &   0&.476 &   0&.686 & 12&  \\
 3679&.639& 3679&.35    & ~\ion{H }{1} & H21     &   0&.550 &   0&.791 & 10&  \\
 3683&.145& 3682&.81    & ~\ion{H }{1} & H20     &   0&.509 &   0&.732 & 12&  \\
 3687&.112& 3686&.83    & ~\ion{H }{1} & H19     &   0&.530 &   0&.760 & 12&  \\
 3691&.901& 3691&.55    & ~\ion{H }{1} & H18     &   0&.596 &   0&.854 & 10&  \\
 3697&.557& 3697&.15    & ~\ion{H }{1} & H17     &   0&.740 &   1&.06  & 10&  \\
 3703&.087& 3702&.75    & ~\ion{O }{3} & 14      &   0&.428 &   0&.611 & 12&  \\
 3704&.220& 3703&.85    & ~\ion{H }{1} & H16     &   0&.799 &   1&.14  &  8&  \\
 3705&.333& 3705&.02    & ~\ion{He}{1} & 25      &   0&.455 &   0&.649 & 12&  \\
 3707&.564& 3707&.27    & ~\ion{O }{3} & 14      &   0&.310 &   0&.442 & 15&  \\
 3712&.369& 3711&.97    & ~\ion{H }{1} & H15     &   0&.964 &   1&.37  &  8&  \\
 3722&.223& 3721&.94    & ~\ion{H }{1} & H14     &   1&.63  &   2&.31  &  6&  \\
 \mcnd    & 3721&.94    & [\ion{S }{3}]& F2      &   \mcnd  &   \mcnd  & \mcnd\\
 3726&.536& 3726&.03    & [\ion{O }{2}]& F1      &  10&.67  &  15&.10  &  2&.5\\
 3729&.284& 3728&.82    & [\ion{O }{2}]& F1      &   5&.63  &   7&.96  &  4&  \\
 3734&.726& 3734&.37    & ~\ion{H }{1} & H13     &   1&.63  &   2&.30  &  6&  \\
 3750&.522& 3750&.15    & ~\ion{H }{1} & H12     &   2&.13  &   3&.00  &  5&  \\
 3755&.083& 3754&.69    & ~\ion{O }{3} & 2       &   0&.630 &   0&.883 & 10&  \\
 3757&.535& 3757&.24    & ~\ion{O }{3} & 2       &   0&.254 &   0&.356 & 15&  \\
 3760&.253& 3759&.87    & ~\ion{O }{3} & 2       &   1&.00  &   1&.40  &  8&  \\
 3771&.005& 3770&.63    & ~\ion{H }{1} & H11     &   2&.62  &   3&.66  &  5&  \\
 3774&.287& 3774&.02    & ~\ion{O }{3} & 2       &   0&.225 &   0&.313 & 15&  \\
 3791&.646& 3791&.27    & ~\ion{O }{3} & 2       &   0&.204 &   0&.283 & 15&  \\
 3798&.271& 3797&.8     & [\ion{S }{3}]& F2      &   3&.66  &   5&.06  &  4&  \\
 \mcnd    & 3797&.92    & ~\ion{H }{1} & H10     &   \mcnd  &   \mcnd  & \mcnd\\
 3820&.013& 3819&.62    & ~\ion{He}{1} & 22      &   0&.741 &   1&.02  &  8&  \\
 3835&.761& 3835&.39    & ~\ion{H }{1} & H9      &   5&.05  &   6&.90  &  3&  \\
 3869&.148& 3868&.75    & [\ion{Ne}{3}]& F1      &  90&.5   & 122&.4    &  1&.2\\
 3889&.230& 3888&.65    & ~\ion{He}{1} & 2       &  16&.4   &  22&.0   &  2&.0\\
 \mcnd    & 3889&.05    & ~\ion{H }{1} & H8      &   \mcnd  &   \mcnd  & \mcnd\\
 3965&.136& 3964&.73    & ~\ion{He}{1} & 5       &   0&.427 &   0&.561 & 10&  \\
 3967&.822& 3967&.46    & [\ion{Ne}{3}]& F1      &  24&.5   &  32&.2   &  1&.5\\
 3970&.464& 3970&.07    & ~\ion{H }{1} & H7      &  11&.94  &  15&.69  &  2&.0\\
 4009&.594& 4009&.26    & ~\ion{He}{1} & 55      &   0&.156 &   0&.203 & 15&  \\
 4026&.593& 4025&.61    & ~\ion{He}{2} & 4.13    &   1&.79  &   2&.32  &  5&  \\
 \mcnd    & 4026&.21    & ~\ion{He}{1} & 21      &   \mcnd  &   \mcnd  & \mcnd\\
 4069&.169& 4068&.60    & [\ion{S }{2}]& F1      &   0&.690 &   0&.884 &  8&  \\
 4070&.325& 4069&.62    & ~\ion{O }{2} & 10      &   0&.151 &   0&.194 & 15&  \\
 \mcnd    & 4069&.89    & ~\ion{O }{2} & 10      &   \mcnd  &   \mcnd  & \mcnd\\
 4072&.560& 4072&.16    & ~\ion{O }{2} & 10      &   0&.100 &   0&.128 & 20&  \\
 4076&.832& 4076&.35    & [\ion{S }{2}]& F1      &   0&.261 &   0&.333 & 12&  \\
 4089&.587& 4089&.29    & ~\ion{O }{2} & 48a     &   0&.086 &   0&.110 & 20&  \\
 4097&.728& 4097&.25    & ~\ion{O }{2} & 20      &   0&.729 &   0&.927 &  8&  \\
 \mcnd    & 4097&.26    & ~\ion{O }{2} & 48b     &   \mcnd  &   \mcnd  & \mcnd\\
 \mcnd    & 4097&.33    & ~\ion{N }{3} & 1       &   \mcnd  &   \mcnd  & \mcnd\\
 4100&.481& 4100&.05    & ~\ion{He}{2} & 4.12    &   0&.166 &   0&.211 & 15&  \\
 4102&.138& 4101&.74    & ~\ion{H }{1} & H6      &  20&.6   &  26&.1   &  1&.5\\
 4103&.736& 4103&.39    & ~\ion{N }{3} & 1       &   0&.500 &   0&.635 & 10&  \\
 4121&.221& 4120&.84    & ~\ion{He}{1} & 16      &   0&.215 &   0&.272 & 15&  \\
 4144&.142& 4143&.76    & ~\ion{He}{1} & 53      &   0&.263 &   0&.331 & 12&  \\
 4169&.542& 4168&.97    & ~\ion{He}{1} & 52      &   0&.032 &   0&.040 & 35&  \\
 \mcnd    & 4169&.22    & ~\ion{O }{2} & 19      &   \mcnd  &   \mcnd  & \mcnd\\
 4200&.297& 4199&.83    & ~\ion{He}{2} & 4.11    &   0&.203 &   0&.251 & 15&  \\
 4227&.960& 4227&.20    & [\ion{Fe}{5}]& F2      &   0&.198 &   0&.243 & 15&  \\
 \mcnd    & 4227&.74    & ~\ion{N }{2} & 33      &   \mcnd  &   \mcnd  & \mcnd\\
 4267&.657& 4267&.15    & ~\ion{C }{2} & 6       &   0&.076 &   0&.092 & 20&  \\
 4339&.163& 4338&.67    & ~\ion{He}{2} & 4.10    &   0&.260 &   0&.310 & 12&  \\
 4340&.888& 4340&.47    & ~\ion{H }{1} & H5      &  39&.6   &  47&.2   &  1&.2\\
 4352&.061& 4351&.81    & [\ion{Fe}{2}]&         &   0&.088 &   0&.104 & 20&  \\
 4363&.640& 4363&.21    & [\ion{O }{3}]& F2      &  13&.19  &  15&.57   &  2&.0\\
 4388&.373& 4387&.93    & ~\ion{He}{1} & 51      &   0&.477 &   0&.558 & 10&  \\
 4393&.240& 4392&.68    & ~\ion{S }{2}?&         &   0&.092 &   0&.107 & 20&  \\
 4434&.975& 4434&.68    & ~\ion{S }{2}?&         &   0&.078 &   0&.090 & 20&  \\
 4438&.168& 4437&.55    & ~\ion{He}{1} & 50      &   0&.080 &   0&.092 & 20&  \\
 4471&.948& 4471&.50    & ~\ion{He}{1} & 14      &   4&.01  &   4&.56  &  3&  \\
 4477&.648& 4477&.47    & ~\ion{C }{1}?&         &   0&.091 &   0&.103 & 20&  \\
 4542&.063& 4541&.59    & ~\ion{He}{2} & 4.9     &   0&.257 &   0&.285 & 12&  \\
 4571&.694& 4571&.10    & ~\ion{Mg}{1}]& 1       &   0&.048 &   0&.053 & 25&  \\
 4610&.683& 4610&.20    & ~\ion{O }{2} & 92e     &   0&.127 &   0&.138 & 15&  \\
 4634&.592& 4634&.14    & ~\ion{N }{3} & 2       &   0&.288 &   0&.310 & 12&  \\
 4639&.387& 4638&.86    & ~\ion{O }{2} & 1       &   0&.066 &   0&.071 & 25&  \\
 4641&.073& 4640&.64    & ~\ion{N }{3} & 2       &   0&.576 &   0&.618 &  8&  \\
 4642&.311& 4641&.81    & ~\ion{O }{2} & 1       &   0&.187 &   0&.201 & 15&  \\
 \mcnd    & 4641&.85    & ~\ion{N }{3} & 2       &   \mcnd  &   \mcnd  & \mcnd\\
 4647&.945& 4647&.42    & ~\ion{C }{3} & 1       &   0&.049 &   0&.052 & 25&  \\
 4649&.767& 4649&.13    & ~\ion{O }{2} & 1       &   0&.201 &   0&.215 & 15&  \\
 4651&.369& 4650&.84    & ~\ion{O }{2} & 1       &   0&.095 &   0&.102 & 20&  \\
 4656&.845& 4656&.39    & ~\ion{Ne}{1} &         &   0&.068 &   0&.073 & 25&  \\
 4658&.771& 4658&.10    & [\ion{Fe}{3}]& F3      &   0&.118 &   0&.126 & 20&  \\
 4662&.025& 4661&.63    & ~\ion{O }{2} & 1       &   0&.114 &   0&.122 & 20&  \\
 4676&.756& 4676&.24    & ~\ion{O }{2} & 1       &   0&.055 &   0&.058 & 25&  \\
 4686&.204& 4685&.69    & ~\ion{He}{2} & 3.4     &   8&.23  &   8&.70  &  2&.5\\
 4703&.902& 4703&.18    & ~\ion{O }{2} &         &   0&.068 &   0&.072 & 25&  \\
 \mcnd    & 4703&.37    & ~\ion{Ar}{2} &         &   \mcnd  &   \mcnd  & \mcnd\\
 4711&.827& 4711&.37    & [\ion{Ar}{4}]& F1      &   3&.54  &   3&.71  &  4&  \\
 4713&.642& 4713&.17    & ~\ion{He}{1} & 12      &   0&.597 &   0&.626 &  8&  \\
 4740&.670& 4740&.17    & [\ion{Ar}{4}]& F1      &   3&.19  &   3&.31  &  4&  \\
 4751&.751& 4751&.34    & ~\ion{O }{2} &         &   0&.126 &   0&.131 & 20&  \\
 4800&.881& 4800&.11    & ~\ion{Ne}{1}?&         &   0&.110 &   0&.112 & 20&  \\
 4859&.816& 4859&.32    & ~\ion{He}{2} & 4.8     &   0&.524 &   0&.524 & 10&  \\
 4861&.803& 4861&.33    & ~\ion{H }{1} & H4      & 100&.0   & 100&.0   &  1&.0\\
 4922&.426& 4921&.93    & ~\ion{He}{1} & 48      &   1&.23  &   1&.21  &  6&  \\
 4931&.795& 4931&.32    & [\ion{O }{3}]& F1      &   0&.160 &   0&.157 & 15&  \\
 4959&.417& 4958&.91    & [\ion{O }{3}]& F1      & 503&.5   & 489&.4   &  0&.8\\
 5007&.348& 5006&.84    & [\ion{O }{3}]& F1      &1527&.    &1462&.    &  0&.8\\
 5016&.178& 5015&.68    & ~\ion{He}{1} & 4       &   1&.56  &   1&.49  &  6&  \\
 5412&.075& 5411&.52    & ~\ion{He}{2} & 4.7     &   0&.879 &   0&.748 &  8&  \\
 5518&.285& 5517&.71    & [\ion{Cl}{3}]& F1      &   0&.488 &   0&.404 & 10&  \\
 5538&.410& 5537&.88    & [\ion{Cl}{3}]& F1      &   0&.461 &   0&.379 & 10&  \\
 5755&.426& 5754&.64    & [\ion{N }{2}]& F3      &   0&.368 &   0&.284 & 12&  \\
 5876&.242& 5875&.67    & ~\ion{He}{1} & 11      &  18&.43  &  13&.74  &  2&.0\\
 6102&.272& 6101&.83    & [\ion{K }{4}]& F1      &   0&.278 &   0&.197 & 15&  \\
 6171&.244& 6170&.70    & ~\ion{He}{2} & 5.18    &   0&.079 &   0&.055 & 25&  \\
 6301&.241& 6300&.30    & [\ion{O }{1}]& F1      &   2&.19  &   1&.49  &  5&  \\
 6312&.256& 6312&.10\TC & [\ion{S }{3}]& F3      &   0&.498 &   0&.338 & 12&  \\
 6312&.892& 6312&.10\TD & [\ion{S }{3}]& F3      &   1&.20  &   0&.814 &  7&  \\
 6364&.728& 6363&.78    & [\ion{O }{1}]& F1      &   0&.865 &   0&.580 &  8&  \\
 6394&.571& 6393&.6     & [\ion{Mn}{5}]&         &   0&.088 &   0&.059 & 25&  \\
 6549&.006& 6548&.03    & [\ion{N }{2}]& F1      &   5&.35  &   3&.46  &  4&  \\
 6560&.754& 6560&.09    & ~\ion{He}{2} & 4.6     &   2&.22  &   1&.44  &  6&  \\
 6563&.442& 6562&.82    & ~\ion{H }{1} & H3      & 456&.3   & 294&.6   &  0&.8\\
 6584&.383& 6583&.41    & [\ion{N }{2}]& F1      &  17&.20  &  11&.07  &  2&.0\\
 6678&.811& 6678&.15    & ~\ion{He}{1} & 46      &   5&.95  &   3&.76  &  4&  \\
 6683&.931& 6683&.21    & ~\ion{He}{2} & 5.13    &   0&.101 &   0&.064 & 25&  \\
 6716&.627& 6716&.47\TC & [\ion{S }{2}]& F2      &   0&.244 &   0&.154 & 15&  \\
 6717&.458& 6716&.47\TD & [\ion{S }{2}]& F2      &   1&.70  &   1&.07  &  6&  \\
 6731&.054& 6730&.85\TC & [\ion{S }{2}]& F2      &   0&.374 &   0&.235 & 12&  \\
 6731&.832& 6730&.85\TD & [\ion{S }{2}]& F2      &   3&.02  &   1&.90  &  5&  \\
 6795&.621& 6795&.00    & [\ion{K }{4}]& F1      &   0&.049 &   0&.031 & 35&  \\
 6891&.566& 6890&.91    & ~\ion{He}{2} & 5.12    &   0&.162 &   0&.099 & 20&  \\
 7006&.347& 7005&.67    & [\ion{Ar}{5}]& 1       &   0&.059 &   0&.036 & 30&  \\
 7065&.946& 7065&.25    & ~\ion{He}{1} & 10      &   6&.62  &   3&.94  &  4&  \\
 7136&.507& 7135&.80    & [\ion{Ar}{3}]& F1      &  13&.2   &   7&.78  &  2&.5\\
 7171&.387& 7170&.62    & [\ion{Ar}{4}]& F2      &   0&.207 &   0&.121 & 20&  \\
 7178&.403& 7177&.53    & ~\ion{He}{2} & 5.11    &   0&.145 &   0&.085 & 20&  \\
 7263&.682& 7262&.76    & [\ion{Ar}{4}]&         &   0&.231 &   0&.134 & 20&  \\
 7282&.072& 7281&.35    & ~\ion{He}{1} & 45      &   1&.10  &   0&.631 &  8&  \\
 7320&.111& 7318&.92    & [\ion{O }{2}]& F2      &   1&.02  &   0&.583 &  8&  \\
 7321&.127& 7319&.99    & [\ion{O }{2}]& F2      &   2&.44  &   1&.40  &  6&  \\
 7330&.702& 7329&.67    & [\ion{O }{2}]& F2      &   1&.56  &   0&.894 &  7&  \\
 7331&.798& 7330&.73    & [\ion{O }{2}]& F2      &   1&.32  &   0&.758 &  8&  \\
 7500&.581& 7499&.82    & ~\ion{He}{1} & 1/8\TE  &   0&.062 &   0&.034 & 35&  \\
 7531&.158& 7530&.54    & [\ion{Cl}{4}]& F1      &   0&.749 &   0&.417 & 10&  \\
 7593&.443& 7592&.75    & ~\ion{He}{2} & 5.10    &   0&.220 &   0&.122 & 20&  \\
 7752&.044& 7751&.12    & [\ion{Ar}{3}]& F1      &   2&.68  &   1&.45  &  6&  \\
 7816&.802& 7816&.16    & ~\ion{He}{1} & 69      &   0&.084 &   0&.045 & 30&  \\
 8046&.477& 8045&.63    & [\ion{Cl}{4}]& F1      &   1&.70  &   0&.888 &  8&  \\
 8237&.562& 8236&.78    & ~\ion{He}{2} & 5.9     &   0&.376 &   0&.193 & 15&  \\
 8242&.450& 8241&.88    & ~\ion{H }{1} & P44     &   0&.080 &   0&.041 & 35&  \\
 8244&.926& 8243&.69    & ~\ion{H }{1} & P43     &   0&.071 &   0&.036 & 35&  \\
 8246&.521& 8245&.64    & ~\ion{H }{1} & P42     &   0&.064 &   0&.033 & 40&  \\
 8248&.598& 8247&.73    & ~\ion{H }{1} & P41     &   0&.084 &   0&.043 & 35&  \\
 8250&.814& 8249&.97    & ~\ion{H }{1} & P40     &   0&.072 &   0&.037 & 35&  \\
 8255&.765& 8255&.02    & ~\ion{H }{1} & P38     &   0&.092 &   0&.047 & 30&  \\
 8258&.704& 8257&.86    & ~\ion{H }{1} & P37     &   0&.098 &   0&.050 & 30&  \\
 8261&.967& 8260&.94    & ~\ion{H }{1} & P36     &   0&.099 &   0&.051 & 30&  \\
 8265&.401& 8264&.29    & ~\ion{H }{1} & P35     &   0&.146 &   0&.075 & 25&  \\
 8268&.943& 8267&.94    & ~\ion{H }{1} & P34     &   0&.131 &   0&.067 & 25&  \\
 8272&.899& 8271&.93    & ~\ion{H }{1} & P33     &   0&.154 &   0&.079 & 25&  \\
 8293&.271& 8292&.31    & ~\ion{H }{1} & P29     &   0&.164 &   0&.084 & 25&  \\
 8307&.046& 8306&.22    & ~\ion{H }{1} & P27     &   0&.228 &   0&.116 & 20&  \\
 8315&.059& 8314&.26    & ~\ion{H }{1} & P26     &   0&.244 &   0&.124 & 20&  \\
 8324&.160& 8323&.43    & ~\ion{H }{1} & P25     &   0&.271 &   0&.138 & 20&  \\
 8334&.647& 8333&.78    & ~\ion{H }{1} & P24     &   0&.238 &   0&.120 & 20&  \\
 8359&.835& 8359&.01    & ~\ion{H }{1} & P22     &   0&.383 &   0&.193 & 20&  \\
 8362&.765& 8361&.77    & ~\ion{He}{1} & 68      &   0&.176 &   0&.089 & 25&  \\
 8375&.278& 8374&.48    & ~\ion{H }{1} & P21     &   0&.356 &   0&.180 & 20&  \\
 8393&.186& 8392&.40    & ~\ion{H }{1} & P20     &   0&.374 &   0&.188 & 20&  \\
 8414&.157& 8413&.32    & ~\ion{H }{1} & P19     &   0&.503 &   0&.253 & 15&  \\
 8438&.771& 8437&.96    & ~\ion{H }{1} & P18     &   0&.590 &   0&.296 & 15&  \\
 8446&.696& 8446&.11    & [\ion{Fe}{2}]& 29      &   0&.088 &   0&.044 & 35&  \\
 8468&.078& 8467&.26    & ~\ion{H }{1} & P17     &   0&.647 &   0&.323 & 15&  \\
 8503&.297& 8502&.49    & ~\ion{H }{1} & P16     &   0&.826 &   0&.410 & 12&  \\
 8665&.816& 8665&.02    & ~\ion{H }{1} & P13     &   1&.59  &   0&.775 & 10&  \\
 8734&.317& 8733&.43    & ~\ion{He}{1} & 6/12\TE &   0&.060 &   0&.029 & 40&  \\
 8751&.323& 8750&.48    & ~\ion{H }{1} & P12     &   2&.04  &   0&.986 & 10&  \\
 8846&.246& 8845&.38    & ~\ion{He}{1} & 6/11\TE &   0&.137 &   0&.066 & 30&  \\
 8863&.646& 8862&.79    & ~\ion{H }{1} & P11     &   2&.77  &   1&.32  &  8&  \\
 8997&.868& 8996&.99    & ~\ion{He}{1} & 6/10\TE &   0&.208 &   0&.098 & 30&  \\
 9015&.747& 9014&.91    & ~\ion{H }{1} & P10     &   3&.63  &   1&.71  &  8&  \\
 9069&.106& 9068&.9\TC  & [\ion{S }{3}]& F1      &   6&.78  &   3&.17  &  6&  \\
 9070&.260& 9068&.9\TD  & [\ion{S }{3}]& F1      &  12&.1   &   5&.66  &  4&  \\
 9211&.235& 9210&.28    & ~\ion{He}{1} & 83      &   0&.224 &   0&.103 & 30&  \\
 9229&.907& 9229&.02    & ~\ion{H }{1} & P9      &   5&.48  &   2&.52  &  7&  \\
 9526&.977& 9526&.17    & ~\ion{He}{1} & 6/8\TE  &   0&.334 &   0&.150 & 30&  \\
 9546&.994& 9545&.98    & ~\ion{H }{1} & P8      &   6&.02  &   2&.69  &  7&  \\
 9560&.951& 9560&.3     & [\ion{Fe}{5}]&         &   0&.129 &   0&.058 & 40&  \\
10028&.61 &10027&.7     & ~\ion{He}{1} & 6/7\TE  &   0&.365 &   0&.157 & 35&  \\
10050&.36 &10049&.4     & ~\ion{H }{1} & P7      &   9&.27  &   3&.98  &  7&  \\
\enddata
\tablenotetext{a}{Where $F$ is the observed flux in units of 
$100.00=1.184\times 10^{-12} {\rm erg}\, {\rm cm}^{-2}\, {\rm s}^{-1}$.}
\tablenotetext{b}{Where $I$ is the reddening corrected flux observed flux,
with $C({\rm H}\beta)=0.59$ dex, in units of 
$100.00=4.606\times 10^{-12} {\rm erg}\, {\rm cm}^{-2}\, {\rm s}^{-1}$.}
\tablenotetext{c}{Blue component.}
\tablenotetext{d}{Red component.}
\tablenotetext{e}{See \citet{peq88}.}
\end{deluxetable}
 
\clearpage

\begin{deluxetable}{l@{\hspace{12pt}}r@{}l@{\hspace{24pt}}c@{\hspace{12pt}}r@{}l}
\tablecaption{Temperatures and Densities
\label{ttemden}}
\tablewidth{0pt}
\tablehead{
\colhead{Lines}  &
\multicolumn{2}{c}{$T_e$ (K)} &
\colhead{Lines} &
\multicolumn{2}{c}{$N_e$(cm$^{-3}$)}}
\startdata
{[\ion{S}{2}]}  & $11230$&$\pm 1100$        & [\ion{S}{2}]  & $5450$&$^{+2650}_{-1550}$ \\
{[\ion{N}{2}]}  & $11180$&$\pm 750$ \TA     & [\ion{O}{2}]  & $2120$&$\pm 240$     \\
{[\ion{O}{2}]}  & $15300$&$\pm 3000$ \TB    & [\ion{Cl}{3}] & $2040$&$^{+1350}_{-1050}$   \\    
{[\ion{S}{3}]}  & $11020$&$\pm 360$         & [\ion{Ar}{4}] & $2920$&$\pm 750$         \\
{[\ion{O}{3}]}  & $11800$&$\pm 100$         \\
$T$(Bac)        & $10700$&$\pm2000$         \\
\enddata
\tablenotetext{a}{Corrected for recombination contribution to the $\lambda$ 5755 auroral 
line, see text.}
\tablenotetext{a}{Corrected for recombination contribution to the $\lambda\lambda$ 
7319,7320,7331, and 7332 auroral lines, see text.}
\end{deluxetable}

\begin{deluxetable}{l@{\hspace{72pt}}r@{}l}
\tablecaption{He$^+$ Ionic Abundance 
\label{thelium}}
\tablewidth{0pt}
\tablehead{
\colhead{Line} & 
\multicolumn{2}{c}{He$^+$/H$^+$\tablenotemark{a}} } 
\startdata
3188    & 7788&$\pm623$ \\ 
3889    & 9389&$\pm376$ \\ 
4026    & 8990&$\pm477$ \\ 
4471    & 8692&$\pm261$ \\ 
4922    & 8965&$\pm538$ \\ 
5876    & 9154&$\pm183$ \\ 
6678    & 9391&$\pm376$ \\ 
7065    & 9090&$\pm364$ \\ 
\hline
Adopted & 9055&$\pm151$\tablenotemark{b} \\ 
\\
\enddata
\tablenotetext{a}{Given in units of $10^{-5}$, for $N_e= 2500$,
$\tau_{3889}=0.93$, and $t^2=0.031$.  Only the line intensity errors are
presented.}
\tablenotetext{b}{In addition to the line intensity errors it also includes the
effects of the uncertainties in $N_e= 2500 \pm 500$, $\tau_{3889}=0.93 \pm
0.57$, and $t^2=0.031\pm0.014$.}
\end{deluxetable}

\clearpage

\begin{deluxetable}{l@{\hspace{48pt}}c}
\tablewidth{0pt}
\tablecaption{He, C, and O Ionic Abundances from Recombination
Lines\tablenotemark{a}
\label{trl}}
\tablehead{
\colhead{Ion} & \colhead{NGC~5307} }
\startdata
He$^+$/H$^+$    & 10.957$\pm$0.007 \\
He$^{++}$/H$^+$ &  9.86$\pm$0.01  \\
C$^{++}$/H$^+$  &  7.95$\pm$0.08  \\
O$^{++}$/H$^+$  &  8.77$\pm$0.03  \\
C$^{+3}$/H$^+$  &  7.46$\pm$0.10  \\
\enddata
\tablenotetext{a}{In units of 12+log(X$^m$/H$^+$).}
\end{deluxetable}

\clearpage

\begin{deluxetable}{l@{\hspace{24pt}}cc}
\tablecaption{Ionic Abundances from Collisionally Excited
Lines\tablenotemark{a}
\label{tcl}}
\tablewidth{0pt}
\tablehead{
\colhead{Ion} & \colhead{$t^2$ = 0.00} & \colhead{$t^2$ = 0.056} }
\startdata
N$^+$      & 6.24$\pm$0.05                 & 6.39$\pm$0.05 \\
O$^+$      & 6.98$\pm$0.08                 & 7.15$\pm$0.08 \\
O$^{++}$   & 8.48$\pm$0.01                 & 8.77$\pm$0.03 \\ 
Ne$^{++}$  & 7.86$\pm$0.03                 & 8.18$\pm$0.04 \\ 
S$^+$      & 5.03$\pm$0.05                 & 5.18$\pm$0.05 \\ 
S$^{++}$   & 6.20$\pm$0.03                 & 6.42$\pm$0.04 \\
Cl$^{++}$  & 4.51$\pm$0.04                 & 4.79$\pm$0.05 \\
Cl$^{+3}$  & 4.72$\pm$0.03                 & 4.96$\pm$0.04 \\ 
Ar$^{++}$  & 5.62$\pm$0.02                 & 5.86$\pm$0.03 \\
Ar$^{+3}$  & 5.61$\pm$0.03                 & 5.91$\pm$0.04 \\
Ar$^{+4}$  & 3.53$\pm$0.14                 & 3.78$\pm$0.14 \\
K$^{+3}$   & 4.13$\pm$0.12                 & 4.40$\pm$0.12 \\
\enddata
\tablenotetext{a}{In units of 12+log(X$^m$/H$^+$).}
\end{deluxetable}

\clearpage

\tabletypesize{\small}
\begin{deluxetable}{ccccccc}
\tablecaption{NGC~5307 Abundance Determinations\tablenotemark{a}
\label{tabund}}
\tablewidth{0pt}
\tablehead{
\colhead{Element}             & \colhead{This paper}           &
\colhead{This paper}          & \colhead{TPP\tablenotemark{b}} &
\colhead{FP\tablenotemark{c}} & \colhead{KB\tablenotemark{d}}  &
\colhead{M\tablenotemark{e}}  \\
                              & \colhead{$t^2 = 0.000$}        &
\colhead{$t^2 = 0.056$}       &\colhead{$t^2 = 0.035$}         &
\colhead{$t^2 = 0.000$}       & \colhead{$t^2 = 0.000$}        &
\colhead{$t^2 = 0.000$}       }
\startdata
He     &   11.00    &   10.99    &   11.01    & 10.94    & $11.02 $ & 10.96 \\
C      &    8.09    &    8.09    &    ....    &  ....    & $ 8.21 $ &  8.43 \\ 
N      &    8.04    &    8.33    &    7.74:   &  8.01:   & $>8.56:$ &  8.16:\\
O      &    8.51    &    8.80    &    8.64    &  8.83    & $ 8.63 $ &  8.55 \\
Ne     &    7.89    &    8.21    &    8.00    &  ....    & $ 7.89 $ &  7.76 \\
S      &    7.02    &    7.24    &    ....    &  6.98    & $ 6.47 $ &  6.81 \\
Cl     &    5.00    &    5.22    &    ....    &  ....    & $ .... $ &  .... \\
Ar     &    5.93    &    6.20    &    ....    &  ....    & $>5.68 $ &  6.14 \\
\enddata
\tablenotetext{a} {In units of $12 +$ Log $N$(X)/$N$(H), gaseous content
only. The C abundances derived in this paper are based on recombination lines,
all the other heavy element abundances on collisionally excited lines.}
\tablenotetext{b} {Torres-Peimbert \& Peimbert (1977).}
\tablenotetext{c} {de Freitas Pacheco et al. (1992).}
\tablenotetext{d} {Kingsburgh \ Barlow (1994).}
\tablenotetext{e} {Mal'kov (1998).}
\end{deluxetable}

\clearpage

\begin{deluxetable}{lr@{}lcccccc@{\hspace{36pt}}l}
\tabletypesize{\footnotesize}
\rotate
\tablecaption{\ion{O}{2} Line Intensity Ratios\tablenotemark{a}
\label{toii}}
\tablewidth{0pt}
\tablehead{ &
\multicolumn{2}{c}{$N_e$(\ion{Cl}{3})} &
\colhead{3p $^4$D$^0_{7/2}$} & 
\colhead{3p $^4$D$^0_{5/2}$} &
\colhead{3p $^4$D$^0_{3/2}$} &
\colhead{3p $^4$D$^0_{1/2}$} &
\colhead{$T_e$(\ion{O}{3})} && \\
\colhead{Object} & 
\multicolumn{2}{c}{(cm$^{-3}$)} &
\colhead{(\tablenotemark{b}\hspace{4pt})} & 
\colhead{(\tablenotemark{c}\hspace{4pt})} & 
\colhead{(\tablenotemark{d}\hspace{4pt})} &
\colhead{(\tablenotemark{e}\hspace{4pt})} & 
\colhead {(K)} &&
\colhead{Source}}
\startdata
NGC~604    & $  100$&$^{+100}_{-50}$         & 0.134 & 0.218 & 0.392 & 0.255 &  8150 && Esteban et al. (2002) \\
30~Doradus & $  295$&$\pm 30$                & 0.158 & 0.254 & 0.372 & 0.216 & 10200 && Peimbert (2003) \\
30~Doradus & $  480$&$\pm 100$               & 0.171 & 0.288 & 0.336 & 0.204 & 10100 && Tsamis et al. (2003) \\
M17        & $ 1050$&$^{+400}_{-400}$        & 0.204 & 0.319 & 0.324 & 0.153 &  8200 && Tsamis et al. (2003) \\
M1-42      & $ 1690$&$^{+600}_{-450}$        & 0.327 & 0.294 & 0.246 & 0.124 &  9220 && Liu et al. (2001) \\
NGC~5307   & $ 2040$&$^{+1350}_{-1050}$      & 0.291 & 0.274 & 0.272 & 0.163 & 11800 && This paper \\
M8         & $ 2400$&$^{+1100}_{-850}$       & 0.237 & 0.263 & 0.317 & 0.183 &  8050 && Esteban et al. (1999b) \\
NGC~3576   & $ 2700$&$\pm 1300$              & 0.240 & 0.250 & 0.241 & 0.269 &  8850 && Tsamis et al. (2003) \\
NGC~7009   & $ 3000$&$\pm 900$               & 0.333 & 0.305 & 0.218 & 0.144 & 10500 && Hyung \& Aller (1995) \\
NGC~6153   & $ 3830$&$\pm 800$ 	             & 0.345 & 0.292 & 0.259 & 0.104 &  9110 && Liu et al. (2000) \\
M2-36      & $ 4830$&$\pm 1000$              & 0.344 & 0.287 & 0.254 & 0.115 &  8380 && Liu et al. (2001) \\
NGC~6543   & $ 5000$&$\pm 1000$              & 0.332 & 0.289 & 0.230 & 0.149 &  8150 && Hyung et al. (2000) \\
Orion~1    & $ 5110$&$\pm 800$ 	             & 0.337 & 0.291 & 0.251 & 0.121 &  8300 && Esteban et al. (1998) \\
Orion~2    & $ 6170$&$\pm 900$               & 0.329 & 0.273 & 0.244 & 0.154 &  8350 && Esteban et al. (1998) \\
NGC~6572   & $20000$&$\pm 4000$              & 0.359 & 0.319 & 0.232 & 0.090 & 10900 && Hyung et al. (1994a) \\
NGC~5315   & $50000$&$^{+50000}_{-25000}$    & 0.390 & 0.302 & 0.214 & 0.093 &  8850 && Peimbert et al. (2003) \\
IC~4997    & \multicolumn{2}{c}{$\sim 10^6$} & 0.352 & 0.282 & 0.216 & 0.150 & 13000 && Hyung et al. (1994b) \\
\\
limit\TF   & \multicolumn{2}{c}{$\infty$}    & 0.397 & 0.300 & 0.202 & 0.101 &  ...  && Wiese et al. (1996) \\
\enddata
\tablenotetext{a}{Intensity ratios of lines belonging to multiplet 1, 3s $^4$P
- 3p $^4$D$^0$. The 8 lines of the multiplet are divided in 4 groups according
to their upper energy level. The upper energy levels are shown in the column
headers.}
\tablenotetext{b}{$I(4649/I$(Sum).}
\tablenotetext{c}{$I(4642 + 4676)/I$(Sum).}
\tablenotetext{d}{$I(4639 + 4662 + 4696)/I$(Sum).}
\tablenotetext{e}{$I(4651 + 4674)/I$(Sum).}
\tablenotetext{f}{High density limit, in this case the line intensities
are proportional to the Einstein A coefficients.}
\end{deluxetable}

\clearpage

\begin{deluxetable}{lr@{$\pm$}lr@{$\pm$}lr@{$\pm$}l}
\tabletypesize{\small}
\tablecaption{NGC~5307, Orion, and Solar Total Abundances\tablenotemark{a}
\label{tta}}
\tablewidth{0pt}
\tablehead{
\colhead{Element}  &
\multicolumn{2}{c}{NGC~5307\tablenotemark{a}} &
\multicolumn{2}{c}{Orion\tablenotemark{b}} & 
\multicolumn{2}{c}{Sun\tablenotemark{c}}}
\startdata
12 + log He/H        & $10.99$ & 0.01    & $10.99$ & 0.01    & $10.98$ & 0.02   \\
12 + log O/H         & $ 8.80$ & 0.03    & $ 8.72$ & 0.06    & $ 8.71$ & 0.05   \\
log C/O              & $-0.71$ & 0.09    & $-0.19$ & 0.08    & $-0.21$ & 0.10   \\
log N/O              & $-0.47$ & 0.15    & $-0.93$ & 0.10    & $-0.78$ & 0.12   \\
log Ne/O             & $-0.59$ & 0.10    & $-0.82$ & 0.12    & $-0.71$ & 0.09   \\
log S/O              & $-1.56$ & 0.12    & $-1.54$ & 0.12    & $-1.51$ & 0.08   \\
log Cl/O             & $-3.58$ & 0.12    & $-3.29$ & 0.13    & $-3.43$ & 0.08   \\
log Ar/O             & $-2.60$ & 0.08    & $-2.23$ & 0.21    & $-2.31$ & 0.08   \\
\enddata
\tablenotetext{a}{Gaseous abundances, values for $t^2$ = 0.056,
obtained in this paper}
\tablenotetext{b}{\citet{pei93,est98,est02b}, values for $t^2$ = 0.024. The O
and C abundances have been increased by 0.08 and 0.10 dex respectively to take
into account the fractions of these elements trapped in dust grains.}
\tablenotetext{c}{\citet{chr98,gre98,all01,all02,hol01}.}
\end{deluxetable}

\clearpage

\begin{figure}
\begin{center}
\includegraphics[scale=0.77]{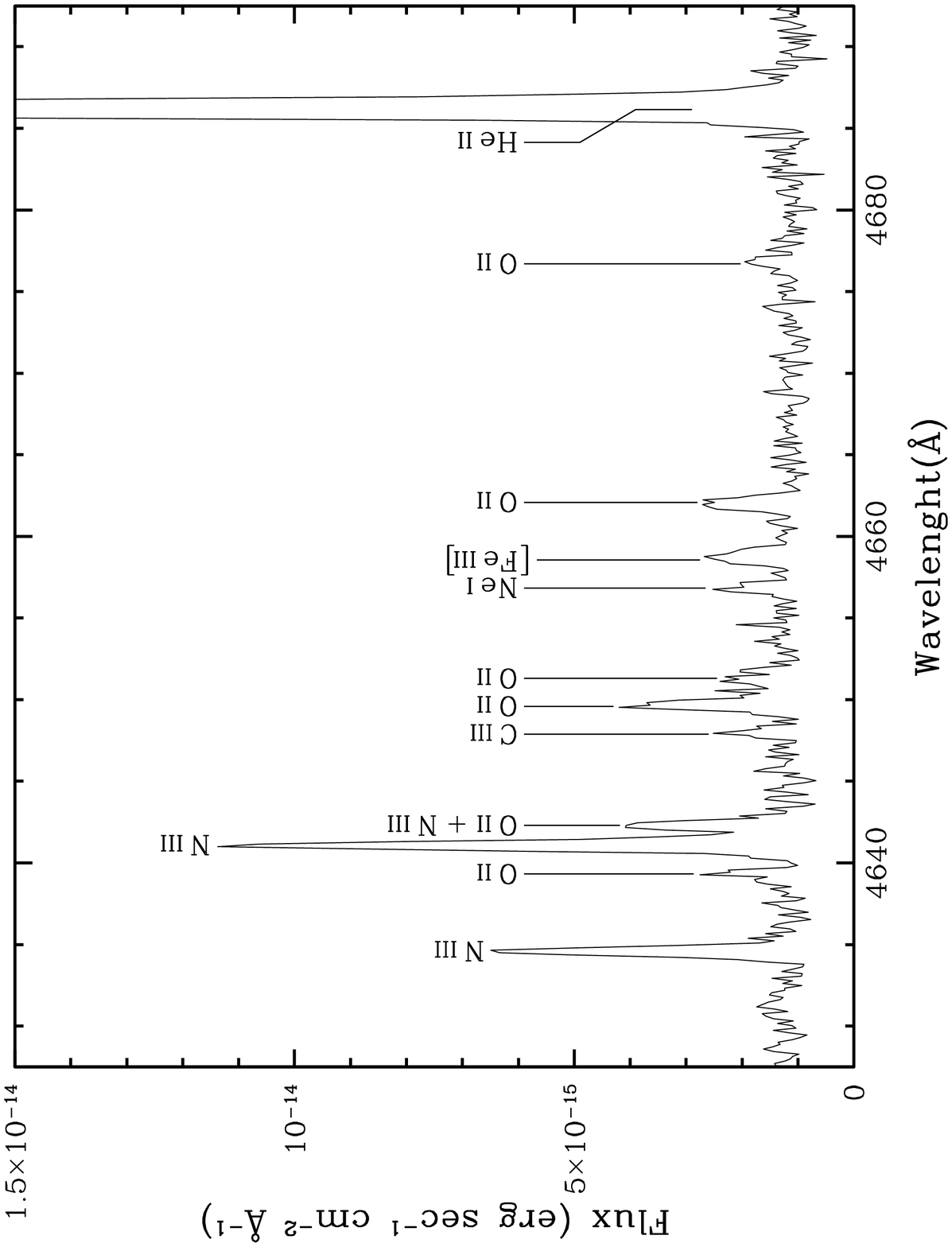}
\end{center}
\end{figure}

\figcaption[f1.eps]{
\label{foxygen}
Section of the echelle spectrum including the \ion{O}{2} multiplet 1 lines
(observed fluxes).  }

\begin{figure}
\begin{center}
\includegraphics[scale=0.77]{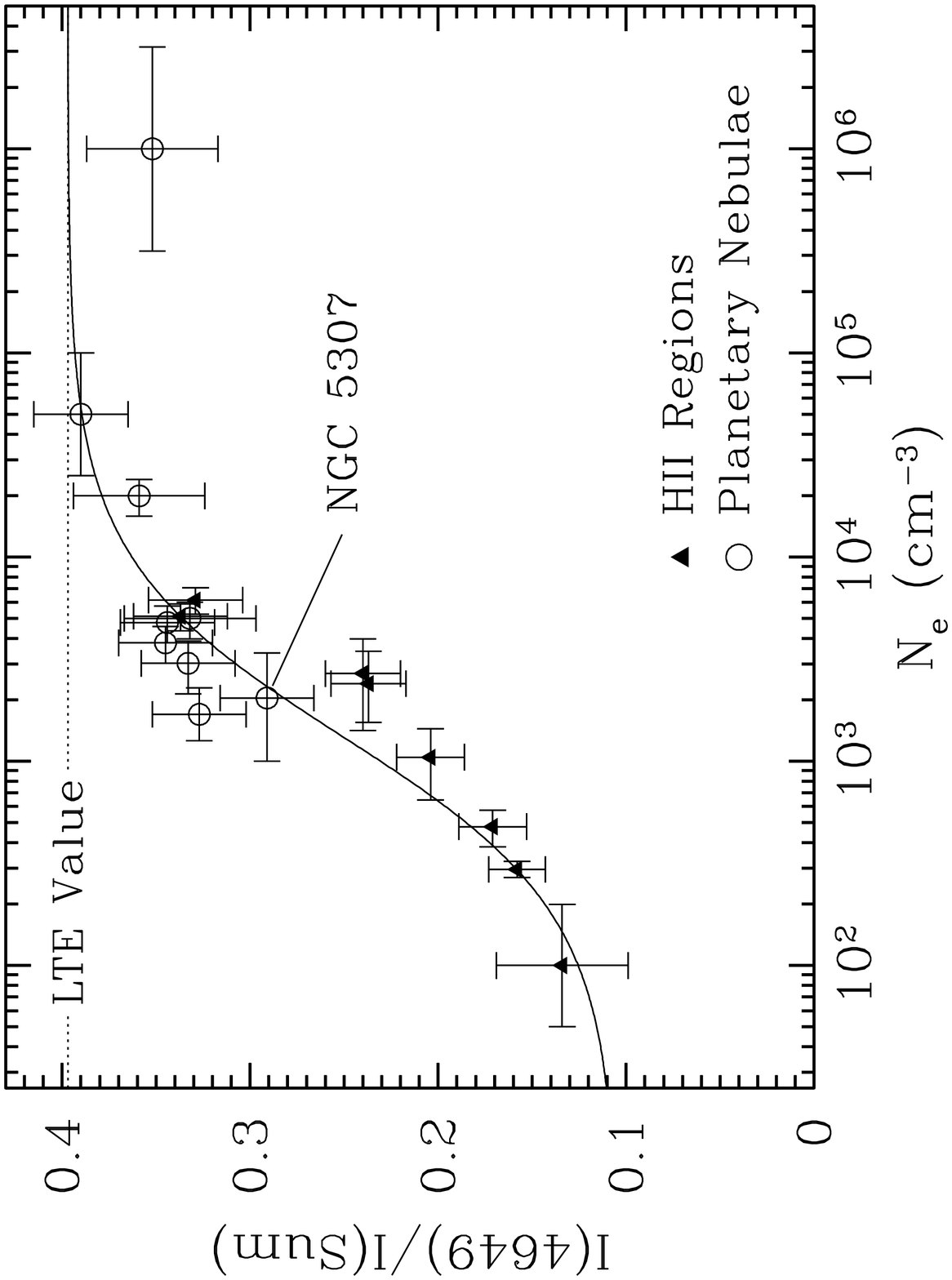}
\end{center}
\end{figure}

\figcaption[f2.eps]{
\label{fOvsden}
Intensity ratio of the \ion{O}{2} multiplet 1 line $\lambda$ 4649 relative to
the sum of the intensities of the eight lines of that multiplet versus the
local electron density derived from forbidden line ratios. The solid line
represents the best fit to the data, see equation~(\ref{ecurve2}) and Table~\ref{toii}.}


\begin{thebibliography}{}

\bibitem[Acker(1976)]{ack76} 
Acker, A. 1976, 
Pub. Obs. Strasbourg 5, fasc. 1

\bibitem[\protect\citeauthoryear{Allende-Prieto, Lambert, \& Asplund}
{Allende-Prieto et~al.}{2001}]{all01} 
Allende-Prieto, C., Lambert, D. L., \& Asplund, M. 2001,
\apj, 556, L63

\bibitem[\protect\citeauthoryear{Allende-Prieto, Lambert, \& Asplund}
{Allende-Prieto et~al.}{2002}]{all02} 
Allende-Prieto, C., Lambert, D. L., \& Asplund, M. 2002,
\apj, 573, L137

\bibitem[Benjamin, Skillman, \& Smits(1999)]{ben99} 
Benjamin, R. A., Skillman, E. D., \& Smits, D. P. 1999,
\apj, 514, 307

\bibitem[\protect\citeauthoryear{Benjamin, Skillman, \& Smits}
{Benjamin et~al.}{2002}]{ben02} 
Benjamin, R. A., Skillman, E. D., \& Smits, D. P. 2002,
\apj, 569, 288

\bibitem[Bowen(1934)]{bow34} 
Bowen, I. S. 1934,
\pasp, 46, 145

\bibitem[Brocklehurst(1971)]{bro71} 
Brocklehurst, M. 1971, 
\mnras, 153, 471

\bibitem[Christensen-Dalsgaard(1998)]{chr98} 
Christensen-Dalsgaard, J. 1998, 
Space Sci. Rev., 85, 19

\bibitem[Davey, Storey, \&  Kisielius(2000)]{dav00} 
Davey, A. R., Storey, P. J., \& 
Kisielius, R. 2000, 
\aaps, 142, 85

\bibitem[de Freitas Pacheco et al.(1992)]{fre92}
de Freitas Pacheco, J. A., Maciel, W. J., \& Costa, R. D. D. 1992,
\aap, 261, 579

\bibitem[D'Odorico et al.(2000)]{dod00}
D'Odorico, S., Cristiani, S., Dekker, H., Hill, V., Kaufer, A.,
Kim, T., \& Primas, F. 2000, 
Proc. SPIE, 4005, 121

\bibitem[Esteban(2002)]{est02a}
Esteban, C. 2002, 
Rev. Mexicana Astron. Astrof\'\i s. Ser. Conf., 12, 56

\bibitem[Esteban et al.(1998)]{est98}
Esteban, C., Peimbert, M., Torres-Peimbert, S., \& Escalante, V.  1998,
\mnras, 295, 401

\bibitem[Esteban et al.(1999a)]{est99a}
Esteban, C., Peimbert, M., Torres-Peimbert, S.,
\& Garc\'{\i}a-Rojas,  J. 1999a,
Rev. Mexicana Astron. Astrof\'\i s., 35, 85

\bibitem[Esteban et al.(1999b)]{est99b} 
Esteban, C., Peimbert, M., Torres-Peimbert, S., Garc\'\i a-Rojas, J., 
\& Rodr\'\i guez, M. 1999b, 
\apjs, 120, 113

\bibitem[Esteban et al.(2002)]{est02b}
Esteban, C., Peimbert, M., Torres-Peimbert, S., \& Rodr\'{\i}guez, M. 2002,
\apj, 581, 241

\bibitem[Ferland(1996)]{fer96}
Ferland, G. J. 1996, 
Hazy, a Brief Introduction to {\sc Cloudy 90} (Univ. Kentucky 
Dept. Phys. Astron. Internal Rep.)

\bibitem[Ferland et al.(1998)]{fer98}
Ferland, G. J., Korista, K. T., Verner,
D. A., Ferguson, J. W., Kingdon, J. B., \& Verner, E. M. 1998, 
\pasp, 110, 761

\bibitem[Garnett(1989)]{gar89} 
Garnett, D. R. 1989, 
\apj, 345, 282

\bibitem[Grandi(1976)]{gra76} 
Grandi, S. A. 1976, 
\apj, 206, 658

\bibitem[Grevesse \& Sauval(1998)]{gre98}
Grevesse, N., \& Sauval, A. J. 1998,
Space Sci. Rev., 85, 161

\bibitem[Guerrero \& Manchado(1996)]{gue96}
Guerrero, M. A., \& Manchado, A. 1996, 
\apj, 472, 711

\bibitem[Holweger(2001)]{hol01}
Holweger, H. 2001,
in Solar and Galactic Composition, ed. R. F. Wimmer-Scweingruber,
AIP Conference Series (New York: Springer), 598, 23

\bibitem[Hyung \& Aller(1995)]{hyu95}
Hyung, S., \&  Aller, L. H. 1995,
\mnras, 273, 973

\bibitem[Hyung et al.(1994a)]{hyu94a}
Hyung, S., Aller, L. H., \& Feibelman, W. A. 1994a,
\mnras, 269, 975

\bibitem[Hyung et al.(1994b)]{hyu94b}
Hyung, S., Aller, L. H., \& Feibelman, W. A. 1994b,
\apjs, 93, 465

\bibitem[Hyung et al.(2000)]{hyu00}
Hyung, S., Aller, L. H., Feibelman, W. A.,
Lee, W. B., \& de Koter, A. 2000,
\mnras, 318, 77

\bibitem[Jacoby \& Ford(1983)]{jac83}
Jacoby, G. H., \& Ford, H. C. 1983, Apj, 266, 298

\bibitem[Kingdon \& Ferland(1995)]{kin95}
Kingdon, J., \& Ferland, G. 1995,
\apj, 442, 714

\bibitem[Kingsburgh \& Barlow(1994)]{kin94}
Kingsburgh, R. L., \& Barlow, M. J. 1994, 
\mnras, 271, 257

\bibitem[Liu (2002)]{liu02}
Liu, X.-W. 2002, 
Rev. Mexicana Astron. Astrof\'\i s. Ser. Conf., 12, 70
 
\bibitem[Liu et al.(2001)]{liu01}
Liu, X.-W., Luo, S.-G., Barlow, M. J., Danziger, I. J., 
\& Storey, P. J. 2001,
\mnras, 327, 141

\bibitem[Liu et al.(2000)]{liu00}
Liu, X.-W., Storey, P. J., Barlow, M. J., Danziger, I. J., Cohen, M.,
\& Bryce, M. 2000,
\mnras, 312, 585

\bibitem[Luridiana et al.(2002)]{lur02}
Luridiana, V., Esteban, C., Peimbert, M. \&  Peimbert, A. 2002,
Rev. Mexicana Astron. Astrof\'\i s., 38, 97

\bibitem[Mal'kov(1998)]{mal98}
Mal'kov, Yu. F. 1998, 
Astronomy Reports, 42, 293

\bibitem[Meatheringham, Wood, \& Faulkner(1988)]{mea88} 
Meatheringham, S. J., Wood, P. R., \& Faulkner, D. J. 1988,
\apj, 334, 862

\bibitem[Mendoza(1983)]{men83}
Mendoza, C. 1983, 
in IAU Symposium 103, Planetary Nebulae, ed. D.R.
Flower (Dordrecht: Reidel), p. 143

\bibitem[O'Dell et al.(2003)]{ode03} 
O'Dell, C. R., Peimbert, M., \& Peimbert, A. 2003, 
\aj, in press

\bibitem[Pagel(1997)]{pag97}
Pagel, B. E. J. 1997, 
Nucleosynthesis and Chemical Evolution of Galaxies, Cambridge:
Cambridge University Press, p. 149

\bibitem[Peimbert(2003)]{pea03}
Peimbert, A. 2003, 
\apj, 584, 735

\bibitem[Peimbert(1967)]{pei67} 
Peimbert, M. 1967, 
\apj, 150, 825

\bibitem[Peimbert(1978)]{pei78}
Peimbert, M., 1978
in IUA Symposium 76, Planetary Nebulae, ed. Y. Terzian
(Dordretch:Reidel), p. 215

\bibitem[Peimbert(1990)]{pei90}
Peimbert, M., 1990
Reports on Progress in Physics, 53, 1559

\bibitem[Peimbert(1993)]{pei93} 
Peimbert, M. 1993, 
Rev. Mexicana Astron. Astrof\'\i s., 27, 9

\bibitem[Peimbert(2002)]{pei02}
Peimbert, M. 2002,  
Rev. Mexicana Astron. Astrof\'\i s. Ser. Conf., 12, 275

\bibitem[Peimbert \& Costero(1969)]{pei69} 
Peimbert, M., \& Costero, R. 1969,
Bol. Obs. To\-nan\-tzin\-tla y Ta\-cu\-ba\-ya, 5, 3

\bibitem[\protect\citeauthoryear{Peimbert, Peimbert, \& Ruiz}
{Peimbert et~al.}{2000}]{pei00}
Peimbert, M., Peimbert, A., \& Ruiz, M. T. 2000,
\apj, 541, 688

\bibitem[Peimbert et al.(2003)]{pei03}
Peimbert, M., Peimbert, A., Ruiz, M. T., \& Esteban, C.  2003,
in preparation

\bibitem[Peimbert, Storey, \& Torres-Peimbert(1993)]{pei93y}
Peimbert, M., Storey, P. J. \& Torres-Peimbert, S. 1993,
\apj, 414, 626

\bibitem[P\'equignot et al.(2002)]{peq02}
Pequignot, D., Amara, M., Liu, X.-W., Barlow, M. J., Storey, P. J., 
Morisset, C., Torres-Peimbert, S. \&  Peimbert, M. 2002
Rev. Mexicana Astron. Astrof\'\i s. Ser. Conf., 12, 142

\bibitem[P\'equignot \& Baluteau(1988)]{peq88}
P\'equignot, D., \& Beluteau, J. P. 1988,
\aap, 206, 298

\bibitem[\protect\citeauthoryear{Rela\~no, Peimbert, \& Beckman}
{Rela\~no et al.}{2002}]{rel02} 
Rela\~no, M., Peimbert, M., \& Beckman, J. 2002, 
\apj, 564, 704

\bibitem[Rubin et al.(2003)]{rub03}
Rubin, R. H., Martin, P. G., Dufour, R. J., Ferland, G. J., Blagrave, 
K.  P. M., Liu, X.-W., Nguyen, J. F., \& Baldwin, J. A. 2003, 
\mnras, in press, (astro-ph/0212244)

\bibitem[Sawey \& Berrington(1993)]{saw93}
Sawey, P. M. J., \& Berrington, K. A., 1993,
Atomic Data and Nuclear Data Tables, 55, 81

\bibitem[Seaton(1979)]{sea79}
Seaton, M. J. 1979,
\mnras, 187, 73p

\bibitem[Smits(1996)]{smi96}
Smits, D. P. 1996,
\mnras, 278, 683

\bibitem[Storey(1994)]{sto94}
Storey, P. J. 1994,
\aap, 282, 999

\bibitem[Storey \& Hummer(1995)]{sto95}
Storey, P. J., \& Hummer, D. G. 1995,
\mnras, 272, 41

\bibitem[Torres-Peimbert \& Peimbert(1971)]{tor71}
Torres-Peimbert, S., \& Peimbert, M. 1971,
Bol. Obs. To\-nan\-tzin\-tla y Ta\-cu\-ba\-ya, 6, 37

\bibitem[Torres-Peimbert \& Peimbert(1977)]{tor77}
Torres-Peimbert, S., \& Peimbert, M. 1977, 
Rev. Mexicana Astron. Astrof\'\i s., 2, 181.

\bibitem[Torres-Peimbert \& Peimbert(2003)]{tor03}
Torres-Peimbert, S., \& Peimbert, M. 2003, 
in Planetary Nebulae and Their Role in the Universe, 
IAU Symposium 209, in press, (astro-ph/0204087)

\bibitem[Torres-Peimbert, Peimbert \& Pe\~na(1990)]{tor90}
Torres-Peimbert, S., Peimbert, M. \&  Pe\~na, M. 1990,
\aap, 233, 540

\bibitem[Tsamis et al.(2003)]{tsa03}
Tsamis, Y. G., Barlow, M. J., Liu, X.-W., Danziger, I. J., 
\& Storey, P. J. 2003,
\mnras, 338, 687

\bibitem[Tylenda et al.(2003)]{tyl03}
Tylenda, R., Siodmak, N., Gorny, S. K., \& Schwarz H. E. 2003,
\aap, in press, (astro-ph/0304433)

\bibitem[Wesson et al.(2003)]{wes03}
Wesson, R., Liu, X.-W., \& Barlow, M. J. 2003,
\mnras, in press, (astro-ph/0301119)

\bibitem[Wiese et al.(1996)]{wie96}
Wiese, W. L., Fuhr, J. R., \& Deters, T. M. 1996,
in Atomic Transition Probabilities of Carbon, Nitrogen, and
Oxygen: A Critical Data Compilation, Journal of Physical and
Chemical Data, Monograph No. 7

\end{thebibliography}
\end{document}